\documentclass[conference, ]{IEEEtran}
\IEEEoverridecommandlockouts
\usepackage{cite}
\usepackage{amsmath,amssymb,amsfonts}
\usepackage{algorithmic}
\usepackage{graphicx}
\usepackage{textcomp}
\usepackage{xcolor}
\usepackage{soul}
\usepackage{hyperref}
\usepackage{multirow}
\usepackage{booktabs}

\usepackage{booktabs} 
\usepackage{array}

\newcommand{\etal}{et al.}
\newcommand{\etals}{et al.'s}

\newcommand{\eg}{{e.g.,}}

\definecolor{cogColor}{HTML}{228B22}
\newcommand{\cdn}[1]{{\textsc{#1}}}

\newcommand{\task}[1]{\textsc{#1}}

\newcommand{\reading}{\task{{reading}}}
\newcommand{\writing}{\task{{writing}}}

\newcommand{\authoring}{\task{{authoring}}}
\newcommand{\modifying}{\task{{modifying}}}
\newcommand{\maintaining}{\task{{maintaining}}}

\newcommand{\taskAlt}[1]{\textsc{{#1}}}
\newcommand{\tbl}{\taskAlt{tabular}}
\newcommand{\cfg}{\taskAlt{config}}
\newcommand{\DSF}{DSF}
\newcommand{\DSFs}{DSFs}
\newcommand{\Nvcrowd}{215}      
\newcommand{\Nrcrowd}{27} 
 
\newcommand{\Nvcrowdp}{4} 
\newcommand{\Ncrowdp}{25}

\newcommand{\Ninterviews}{9}    
 
\newcommand{\pdavid}{P\textsubscript{01}} 
\newcommand{\pyu}{P\textsubscript{02}}
\newcommand{\pdominik}{P\textsubscript{03}}
\newcommand{\pallen}{P\textsubscript{04}} 
\newcommand{\prahat}{P\textsubscript{05}}
\newcommand{\pdaniel}{P\textsubscript{06}}
\newcommand{\pmorris}{P\textsubscript{07}}

\newcommand{\pstun}{P\textsubscript{09}}

\definecolor{steelblue}{RGB}{70, 130, 180}

\hypersetup{
    colorlinks=true, 
    linkcolor=steelblue, 
    urlcolor=steelblue,
    citecolor=steelblue
}
\let\oldhref\href

\newcommand{\ourhref}[2]{\oldhref{#1}{\textcolor{steelblue}{$\nearrow$ #2}}}

\newcommand{\hlc}[2][yellow]{{%
                  \colorlet{foo}{#1}%
                  \sethlcolor{foo}\hl{#2}}%
}
\definecolor{hnColor}{HTML}{ff6600}
\definecolor{inColor}{HTML}{9ed0ff}
\newcommand\hn[1]{\hlc[hnColor!40]{``#1''}}

\newcommand\interview[1]{\hlc[inColor!40]{``#1''}}

\newcommand{\secref}[1]{\hyperref[#1]{Sec.~\ref*{#1}}}
\newcommand{\appendixref}[1]{\hyperref[#1]{Appendix~\ref*{#1}}}
\newcommand{\figref}[1]{\hyperref[#1]{Fig.~\ref*{#1}}}
\newcommand{\eqnref}[1]{\hyperref[#1]{Eqn.~\ref*{#1}}}
\newcommand{\tabref}[1]{\hyperref[#1]{Table ~\ref*{#1}}}

\def\subsubsec#1
{\subsubsection{#1}}

\definecolor{linkColor}{HTML}{257E98}

\newcommand{\osf}{\url{https://osf.io/jryna/}}

\newcommand{\paraheadd}[1]
{%
    \vspace{0.07in}%
    \noindent%
    \textbf{\textit{#1}}%
}
\newcommand{\parahead}[1]{\paraheadd{#1.}}
\def\subsubsec#1
{\subsubsection{#1}}

\def\BibTeX{{\rm B\kern-.05em{\sc i\kern-.025em b}\kern-.08em
    T\kern-.1667em\lower.7ex\hbox{E}\kern-.125emX}}
\begin{document}

\title{Reading Between the Curly Braces:\\ On Textual Data Serialization Format Usability 
}

\author{\IEEEauthorblockN{1\textsuperscript{st} Shiyi He}
\IEEEauthorblockA{\textit{University of Utah} \\
Salt Lake City, Utah 
}
\and
\IEEEauthorblockN{2\textsuperscript{nd} Zach Cutler}
\IEEEauthorblockA{\textit{University of Utah} \\
Salt Lake City, Utah 
}
\and
\IEEEauthorblockN{3\textsuperscript{rd} Andrew M. McNutt}
\IEEEauthorblockA{\textit{University of Utah} \\
Salt Lake City, Utah 
}
}


\maketitle

\begin{abstract}
Textual data serialization formats, such as JSON or XML, are ubiquitous, supporting tasks like software configuration and data tabularization. Despite their prominence, little is known about their usability. What makes one good or bad? Is there a best one for cognitive efficiency?  We explore these questions via a ($N=\Nvcrowd{}$) crowd work study and a ($N=\Ninterviews{}$) semi-structured interview study. 
We find that format distinctions (like indentation versus curly braces) do not consistently translate into substantial usability differences.
While HJSON and YAML performed better than other formats in certain modification tasks, these advantages disappeared in more realistic settings where task complexity was either trivial or highly demanding. 
Instead, usability appears driven by sociotechnical ecosystems: the tooling, documentation, and community practices surrounding a format matter more than syntax.

\end{abstract}


\section{Introduction}

Textual data serialization formats (\DSFs{}) such as JSON and XML are ubiquitous parts of modern computing.
They are used for a constellation of different tasks and are routinely read, edited, validated, and debugged by humans in workflows including configuring software, tabulating data, maintaining website data files, and specifying programs in small domain-specific languages~\cite{mcNutt2022grammar}.
This makes their human usability consequential across both professional and everyday technical work.
Debates on the merits of the usability of various \DSFs{} are common and polarizing, frequently centering on readability and writability.
For instance, some developers enjoy JSON \hn{because it's lightweight and easy for humans to read and write.}~\cite{cizhu}
Others argue the opposite, suggesting that \hn{JSON just isn't very good for human reading/writing. It's just too strict and syntax-heavy.}~\cite{deleted}\footnote{We quote social media accounts \hn{like so} and interviewees \interview{like so}, a convention that supports our interview study, \secref{sec:rq2}.}
Similar disagreements appear elsewhere.
For instance, Mahmoudi \etal{}~\cite{mahmoudi2024from} recount a debate in choosing the host syntax for GeoJSON, saying ``an extended argument over XML and JSON ensued,'' which was characterized as a ``religious-war''.
New \DSFs{} continue to emerge (\eg{} CSON~\cite{cson},  CUE~\cite{cuelang}, DMS~\cite{dms}, EDN~\cite{edn}, KDL~\cite{kdl}, LEAN~\cite{lean-format}, MAML~\cite{maml}, PKL~\cite{pkl}, KSON~\cite{kson}, RON~\cite{ron}, TOON~\cite{toon}, TRON~\cite{tron}, and ZON~\cite{zon}), indicating dissatisfaction with existing formats, and a belief that usability issues can be resolved via notational design variations.

Despite the strength of these opinions, the semantics of \DSFs{} are essentially identical---in most cases each of these formats can be losslessly transformed from one to another.
Yet, prior work~\cite{meyerovich2013empirical,meyerovich2012socio} suggests that factors such as open source libraries, existing code, and experience more significantly impact programming language adoption, compared to syntactic elements.
This raises a compelling question: \emph{do similar dynamics govern \DSF{} usability? }

Prior work~\cite{nurseitov2009comparison, sumaray2012comparison, maeda2012performance,liu2025data} on \DSFs{} primarily focuses on machine-oriented metrics such as parsing speed or storage efficiency, leaving human factors largely unexplored.
As a result, ongoing discussions about \DSFs{} usability rely on intuition and anecdote rather than empirical findings. The demand for such evidence is visible in community discourse. For example, while discussing curly braces versus indentation in \DSFs{}, one poster asked \hn{are there any case studies that show [indentation is] more readable? I'm happy to accept that it is, but I can't help wondering if research has been done, or it's mostly gut feeling/anecdotes/aesthetics.}~\cite{diob}.
We take up this challenge via an exploration of \DSF{} usability, through two research questions.

\vspace{-0.4em}

\paraheadd{To what extent are differences in \DSFs{} usability attributable to notational design?}
We conducted a crowd-work study ($N=\Nvcrowd{}$) examining performance on everyday reading and writing tasks with different textual \DSFs{}.
If notational design plays a dominant role, we would expect to observe systematic performance differences between \DSFs{} across tasks.
We observe no systematic performance advantage for any single format, aside from some minor differences in certain editing tasks.
We note that usability variance diminishes at task complexity extremes: as the task approaches either trivial or increasingly difficult, usability differences become less noticeable.

\begin{figure*}
    \centering
    \includegraphics[width=\linewidth]{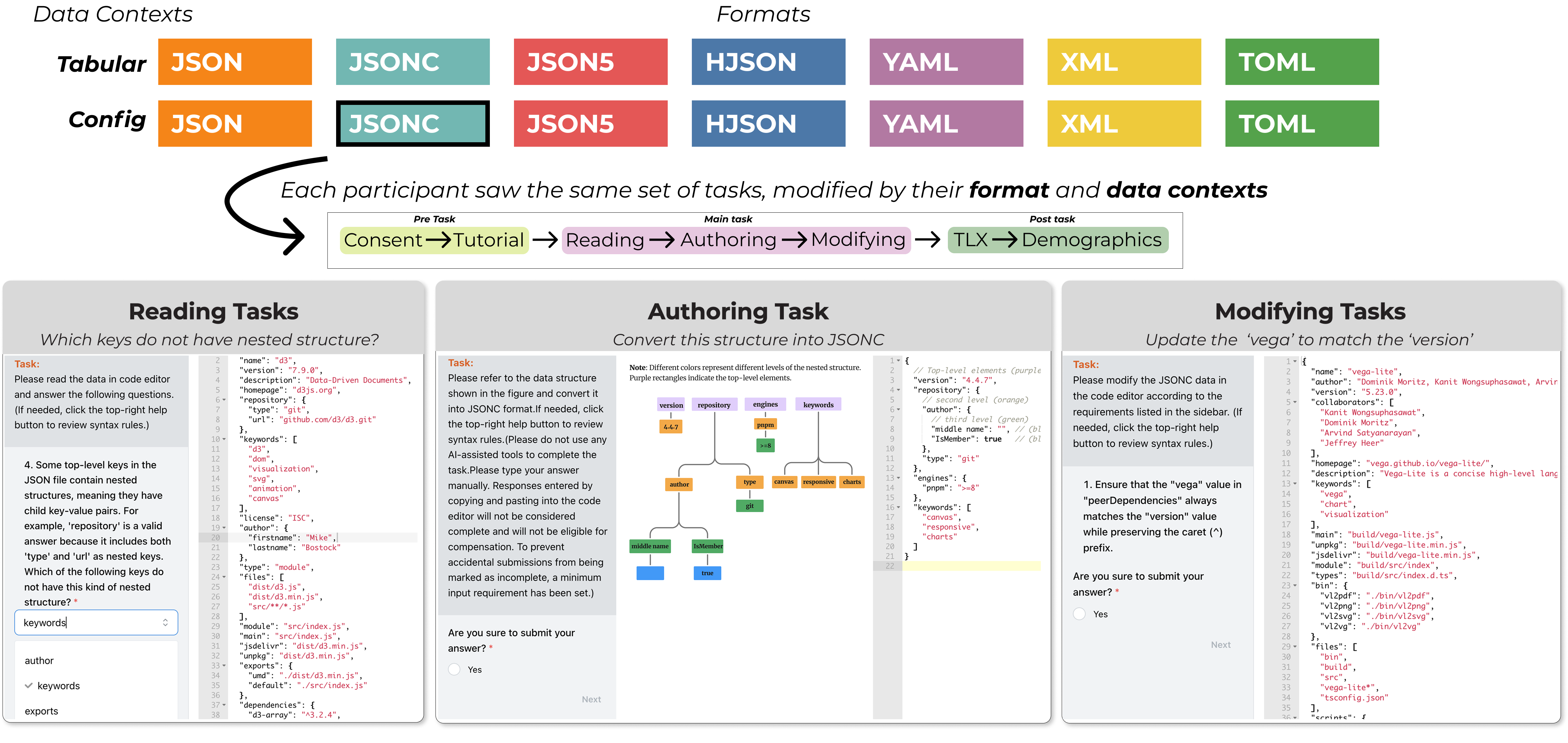}
    \vspace{-2em}
    \caption{Stimuli used in crowd-work experiment for JSONC in the \cfg{} task.
        An example of the study can be found \ourhref{https://dsf-study.netlify.app/a-usability-study}{here}.
    }
    \label{fig:stimuli-examples}
    \vspace{-1em}
\end{figure*}

\vspace{-0.4em}
\paraheadd{What are the factors that drive usability beyond notational design?}
We conducted
a ($N=\Ninterviews{}$) semi-structured interview study with practitioners, which surfaces real-world experiences, challenges, and decision-making processes surrounding \DSF{} use.
Our findings indicate that \DSF{} usability is strongly shaped by sociotechnical and human factors---including community norms, tooling support, ecosystem integration, and individual preferences---echoing prior work on programming language adoption~\cite{meyerovich2013empirical,meyerovich2012socio}.

\paraheadd{}Between these studies
we find that notational differences do not have robust or consistent effects on usability across common data tasks, with no evidence of a single ``best'' format. Instead, usability is primarily shaped by sociotechnical and human factors, including community conventions, tooling ecosystems, and individual aesthetic preferences. These findings reframe existing debates about \DSFs{} by shifting attention away from syntax comparisons toward a sociotechnical and human-centered understanding of these formats.

\section{Related work}

Our work focuses on human-readable \DSFs{} that express content in plain text, which we position in contrast to those that use binary representations (\eg{} Avro~\cite{avroSpec}, Thrift~\cite{thriftSpec}, and Apache Arrow~\cite{arrowSpec}).
Similarly, unlike prior evaluations of \DSFs{}~\cite{nurseitov2009comparison, sumaray2012comparison, maeda2012performance, liu2025data}, which emphasize machine-oriented metrics like serialization efficiency, we focus on human usability.

\parahead{Data Serialization Formats}
Human-readable \DSFs{} have evolved considerably over time, moving from simple flat encodings toward more expressive hierarchical structures to better support human readability, editing, and interoperability.
Early formats such as CSV (1970s)~\cite{csvSpec} (and its descendants such as TSV from the 1990s~\cite{tsvSpec}) provided flat tabular data using simple delimiters. Between these INI (1980s)~\cite{iniSpec} adopted a simple key-value syntax suitable for some configuration tasks. These formats were easy to parse but limited in their ability to represent nested or complex data.

As the web expanded in the mid-1990s, the need for structured and interoperable data formats grew~\cite{gillies2000web}.  XML~\cite{xmlSpec} was introduced to address these challenges, featuring a verbose tag-based syntax (\eg{} \verb+<body></body>+) and building on earlier standards such as HTML~\cite{htmlSpec}. It quickly became foundational in web and enterprise ecosystems, supported by a broad tooling infrastructure (\eg{} XSLT~\cite{clark1999xslt}  and XML Schema~\cite{thompson2001xmlschema1}). However, its verbosity and complexity later motivated the development of lighter-weight alternatives.

In the early 2000s, JSON~\cite{crockford_json_org} emerged as a simpler, JavaScript-aligned format for hierarchical data and was quickly adopted especially for web data interchange~\cite{severance2012discovering}. Yet, like its namesake (JavaScript), JSON too was seen as having a variety of ``bad parts''~\cite{crockford2008javascript}---such as readability issues, a lack of comments, and strict syntactic requirements.  Subsequent formats, such as YAML~\cite{yamlSpec} and TOML~\cite{tomlSpec}, aimed to improve readability and developer ergonomics by reducing syntactic noise and introducing alternative indentation for nesting. Variants such as JSONC~\cite{jsoncSpec}, HJSON~\cite{hjsonSpec} and JSON5~\cite{json5Spec} further relaxed strict syntax rules to improve flexibility.  A broader range of formats have been explored for various purposes (\eg{} RDF~\cite{rdf11concepts}, HOCON~\cite{hocon}, SDlang~\cite{sdlang}) further reflects ongoing experimentation in this space.
Beyond these, a range of other formats~\cite{cson, edn, kdl, cuelang, ron, pkl, kson, tron, zon, maml} have been developed to support varied aesthetic preferences or contextual constraints---for instance TOON~\cite{toon} focuses on minimizing LLM tokens.
This proliferation highlights the need to better understand the usability of the domain.
Complementing these formats are a constellation of auxiliary tools that mediate their use, such as syntax highlighters, autoformatters, and linters.
While additional tools exist and pose important usability questions, our focus is on the usability of the notation itself rather than a particular tool or implementation.

\parahead{Human-Centered Perspectives on Language-like Artifacts}
A key motivation for this work is the limited understanding of human-centered usability of \DSFs{}.
Sumaray \etal{}~\cite{sumaray2012comparison} briefly speculate on the usability of \DSFs{} and note that usability in this domain remains difficult to quantify.
Gryk~\cite{gryk2022human} reflects on the human readability of data files, proposing a metric for human readability of structured data---akin to structures like Flesch-Kincaide~\cite{kincaid1975derivation} readability scores.
Lassila \etal{}~\cite{lassila2015comparison} find that syntactic differences between  XML query languages meaningfully affected learnability.
Graaumans~\cite{graaumans2004a} found that concise, cognitively aligned languages like XQuery lead to faster task completion and fewer errors.
Our work complements these by considering both the social context and low-level usability of the wider class of \DSFs{}.

Meyerovich and Rabkin~\cite{meyerovich2013empirical,meyerovich2012socio} analyze usage and survey data to understand how practitioners adopt different programming languages.
They find that intrinsic language features have only secondary importance in adoption; open-source libraries, existing code, and prior experience strongly influence language selection, while features like performance and simple semantics do not.
However, isolating the specific impact of syntax in general-purpose programming languages is difficult due to their functional complexity and heterogeneous usage contexts---an issue which is neatly resolved by the simplicity of \DSFs{}.
Alongside adoption research, there has been increasing attention to the human-centered usability of programming languages and related notations ~\cite{chasins2021pl, myers2016usability, churchbrief}, with studies examining readability, learnability, debugging practices, and user perception across languages and domain-specific systems \cite{oliveira2020evaluating,dantas2021readability,gamboa2021user,gathani2020debugging}.
Work on usability evaluation for DSLs has proposed structured criteria for assessing effectiveness, efficiency, and user satisfaction \cite{poltronieri2017usability,albuquerque2015quantifying,poltronieri2018usability}. While these approaches provide valuable methodological foundations, they do not directly address \DSFs{}.
Most closely related to our work are Stefik \etals{}~\cite{stefik2013empirical} collection of empirical studies on novice usability and learnability of general purpose programming notations. They find that notational choices \emph{do have an effect} on learnability.
In contrast, we do not directly examine learnability but focus on performance in common data tasks, where notational effects appear minimal.

\begin{figure*}[t]
    \centering
    \footnotesize
    \begin{tabular}{|p{0.47\linewidth}|p{0.47\linewidth}|}
        \hline
        \tbl{}                                                                                                                                                       & \cfg{}                                                                                                                                                                                                                                                                                          \\
        \hline
        \multicolumn{2}{|l|}{\reading{} (\tbl{} \cite{b2t2_example_tables}, \cfg{} \cite{d3_package_json} )}                                                                                                                                                                                                                                                                                                                                                           \\
        \hline
        1. What is the final exam grade for Bob?                                                                                                                     & 1. What is the value of the ``version'' key?                                                                                                                                                                                                                                                    \\[3pt]
        2. Which quiz did Alice achieve the highest grade on?                                                                                                        & 2. What is the value of the ``node'' key?                                                                                                                                                                                                                                                       \\[3pt]
        3. How many quizzes has Bob taken in total, including the midterm and final exam?                                                                            & 3. How many values does the ``keywords'' key contain?                                                                                                                                                                                                                                           \\[3pt]
        4. Which student participates in two different sports?                                                                                                       & 4. Some top-level keys in the JSONC file contain nested structures, meaning they have child key-value pairs. For example, ``repository'' is a valid answer because it includes both ``type'' and ``url'' as nested keys. Which of the following keys do not have this kind of nested structure? \\[3pt]
        5. Which students have not been absent from any quizzes, including the midterm and final exams? List all that apply.                                         & 5. In the ``dependencies'' section, any dependency with a version number of 5.0.0 or higher may potentially cause compatibility issues. Which of the following dependencies meet this criterion?                                                                                                \\
        \hline
        \multicolumn{2}{|l|}{\authoring{} (\tbl{} \cite{patientdata}, \cfg{} \cite{chartjs_package_json})}                                                                                                                                                                                                                                                                                                                                                             \\
        \hline
        Please refer to the data structure shown in the figure and convert it into JSONC format.                                                                     & Please refer to the data structure shown in the figure and convert it into JSONC format                                                                                                                                                                                                         \\
        \hline
        \multicolumn{2}{|l|}{\modifying{} (\tbl{} \cite{wikipedia_movie_data}, \cfg{}  \cite{vegalite_package_json})}                                                                                                                                                                                                                                                                                                                                                  \\
        \hline
        1. Go through each movie entry in the JSONC data and update the ``writer'' element to match the ``director'' element, keeping them consistent.               & 1. Ensure that the ``vega'' value in ``peerDependencies'' always matches the ``version'' value while preserving the caret (\textasciicircum) prefix.                                                                                                                                            \\[3pt]
        2. Check the genres list in each movie entry. If ``Silent'' is not already present, add it; otherwise, leave the list unchanged.                             & 2. Add the keys ``d3-hierarchy'' with value ``\textasciicircum3.1.2'', ``d3-interpolate'' with value ``\textasciicircum3.0.1'', and ``d3-path'' with value ``\textasciicircum3.1.0'' under the ``dependencies'' section.                                                                        \\[3pt]
        3. Remove movie objects where the release date is not from the year 1900.                                                                                    & 3. Go through the JSONC data and remove all ``url'' key-value pairs from the objects, no matter where they are.                                                                                                                                                                                 \\[3pt]
        4. For each movie object in the JSONC file, create a new key called information. Move all existing key-value pairs under this key, except for the title key. & 4. Modify the JSONC structure by transforming the ``collaborators'' list into an array of objects, where each collaborator is represented as an object with a ``name'' key. Each original collaborator string should become the value of a ``name'' key within its respective object.           \\[3pt]
        \hline
    \end{tabular}
    \caption{Tasks used in the crowd-work study (with JSONC as an example).
        The data used were adapted from publicly available datasets.
        These simplified versions preserve representative structural features (\eg{} nested objects and arrays) while reducing complexity to limit the already long experiment scope.
    }
    \label{fig:task-table}
    \vspace{-1em}
\end{figure*}

\section{Setting}
In these studies, we focus on three common interactions users have with \DSFs{}: \reading{}, \writing{} (including \authoring{} and \modifying{}), and \maintaining{}. Our crowd-work user study centers on the former two, while the interview study focuses on \maintaining{} as it unfolds over longer timescales and is difficult to capture through short-term tasks, making it better suited to practitioner accounts.
In addition to these low-level tasks, we considered high-level task contexts. \DSFs{} are commonly used in \cfg{} files (such as manipulating Docker configuration files), \tbl{} data (as in exchanging datasets for analysis or logging purposes). While \DSFs{} are used to in some DSLs (\eg{} Vega-Lite~\cite{satyanarayan2016vega}), we forgo their analysis
as each one comes with its own usability concerns that are closely tied to its specific language design.

We consider five widely used \DSFs{}---CSV, XML, JSON, YAML using only common to YAML 1.1 and 1.2), and TOML---and three JSON variants---JSONC, JSON5, HJSON---based on their syntactic variety, relative popularity, and availability of public facing commentary.
We excluded formats with relatively low visibility and usage (\eg{} EDN, Dhall, and CSON) due to their sparse discussion on technical forums and limited presence in public code repositories. While investigating all \DSFs{} would be preferable, such comparison is out of scope for our human-centered experimental approach.
Focusing on these relatively well-known formats gives a richer ecological validity to our designs.

Finally, an alternative approach would involve varying individual notations to see their individual effects. We instead focused on real formats to center ecologically over construct validity, so as to make our results more transferable.

\section{crowd-work study}
\label{sec:rq1}

Here we consider the extent to which differences in \DSFs{} usability are attributable to notational design. We do so by investigating the usability variances of \DSFs{} across common \reading{} and \writing{} tasks in a crowd-work study.

\subsection{Methodology}
\label{sec:userstudydesign}

Our study was a between-subjects design with data context and \DSF{} factors.

\parahead{Task Design}
All participants followed the experiment sequence shown in ~\figref{fig:stimuli-examples}. Each participant was sorted into one of seven \DSFs{}
and one of two data contexts (\ourhref{https://dsf-study.netlify.app/a-usability-study-config/}{\cfg{}} or \ourhref{https://dsf-study.netlify.app/a-usability-study-tabular/}{\tbl{}}), balanced across both factors.
A warm-up (consisting of a tutorial, practice exercise, typing speed measurement) was followed by three main tasks---\reading{}, \authoring{}, and \modifying{}.
This task design (see \figref{fig:task-table}) was grounded in common interactions that developers perform with \DSFs{} in practice.
In \cfg{} files, users routinely read existing files to extract information, author new structured content, and modify entries to update or correct values.
In \tbl{} files, users commonly inspect rows and columns to identify relevant records, revise cells to fix inconsistencies or computed fields, create new records or reorganize column structures.
We ordered these tasks to reflect a progression of cognitive engagement.
The datasets used in our study were derived from \tbl{} and \cfg{} files found in public GitHub repositories to provide ecologically motivated examples. We selected files from real projects that exhibited structures relevant to our study tasks, such as nesting and mixed value types, using well-known repositories when possible, such as \texttt{package.json} from D3.
The \DSFs{} included in the user study were selected for their comparable representational capacity for the target tasks.
For example, CSV was excluded due to its incompatibility with nested data structures (but was still examined CSV in our interviews). Although XML is not typically used for tabular data storage, we included it as it can represent such structures, enabling comparison of how syntactic differences influence user experience and task performance across \DSFs{}. Task files were formatted (in terms of order and structure) consistently across DSFs while maintaining idiomatic format usage. For additional details, please see the stimuli such as (\ourhref{https://dsf-study.netlify.app/a-usability-study/reviewer-reading-task-config-xml}{XML} or \ourhref{https://dsf-study.netlify.app/a-usability-study/reviewer-reading-task-config-json}{JSON}).

We used the Ace code editor~\cite{ace-editor} with
features like linters and auto-formatting disabled, retaining only syntax highlighting and basic line indentation, to minimize the effect of the editor's affordances on the results.
This study was implemented in reVISit~\cite{cutler25revisit2}, which we used to track timing, keyboard input (\eg{} for Cmd+f usage), and correctness.

\parahead{Participants}
We recruited ($N=\Nvcrowd{}$) participants from Prolific~\cite{prolific}, requiring English proficiency and an approval rate $\geq 95\%$.
Participants were paid approximately \$8.00/hr based on real time taken, with tasks taking 64 (\cfg{}) and 57 (\tbl{}) minutes on average.
We did not require prior programming experience (echoing prior studies~\cite{stefik2013empirical} of programming language design) to avoid bias stemming from prior exposure (rather than differences in the formats themselves).
This yielded a range of novice-leaning but varied experience distribution (\figref{fig:experience-heat}), allowing us to view the effects across experience. However, we did not observe a systematic relationship between prior experience and task performance.

To determine the sample size, we conducted an a priori power analysis using a one-way fixed-effects F-test for ANOVA with 14 groups (7 formats × 2 data contexts) targeting a moderate effect size ($f = 0.3$), $\alpha = 0.05$, and power ($1 - \beta = 0.8$). This suggested $\sim15$ participants per condition would be sufficient to detect meaningful differences, with real sampling yielding 15-17 due to an implementation error.
We iteratively piloted with ($N=\Ncrowdp{}$) Prolific participants, \Nvcrowdp{} of whom completed trials that were the same as the final protocol and so were retained in the main study.
Some ($\Nrcrowd{}$) participants' data was dropped for excessive use of AI (which violated our consent terms), such as pasting large text blocks. As AI detection is error-prone~\cite{weber2023testing}, participants suspected of AI usage were compensated.
This study was marked exempt by the University of Utah IRB.

\parahead{Evaluation}
Reading tasks were evaluated via straightforward scoring of multiple choice questions. Scores for \authoring{} and \modifying{} were assessed through manual grading. Two of the authors met to develop a grading rubric (based on semantic and syntactic correctness), after which both authors separately graded 20 trials---following common practices~\cite{vitak_24_reviewer} inter-rater reliability was not calculated.
They then met to resolve the differences, after which the first author graded the remaining participants.
We developed an autograder to sanity check the data during collection based on parsability and tree-edit distance to prepared solutions---although these values were strongly dependent on parser quality, which was uneven for these formats.
Our goal was to evaluate notations rather than implementations, so we relied on manual grading.

\begin{figure*}[t]
    \centering
    \includegraphics[width=\linewidth]{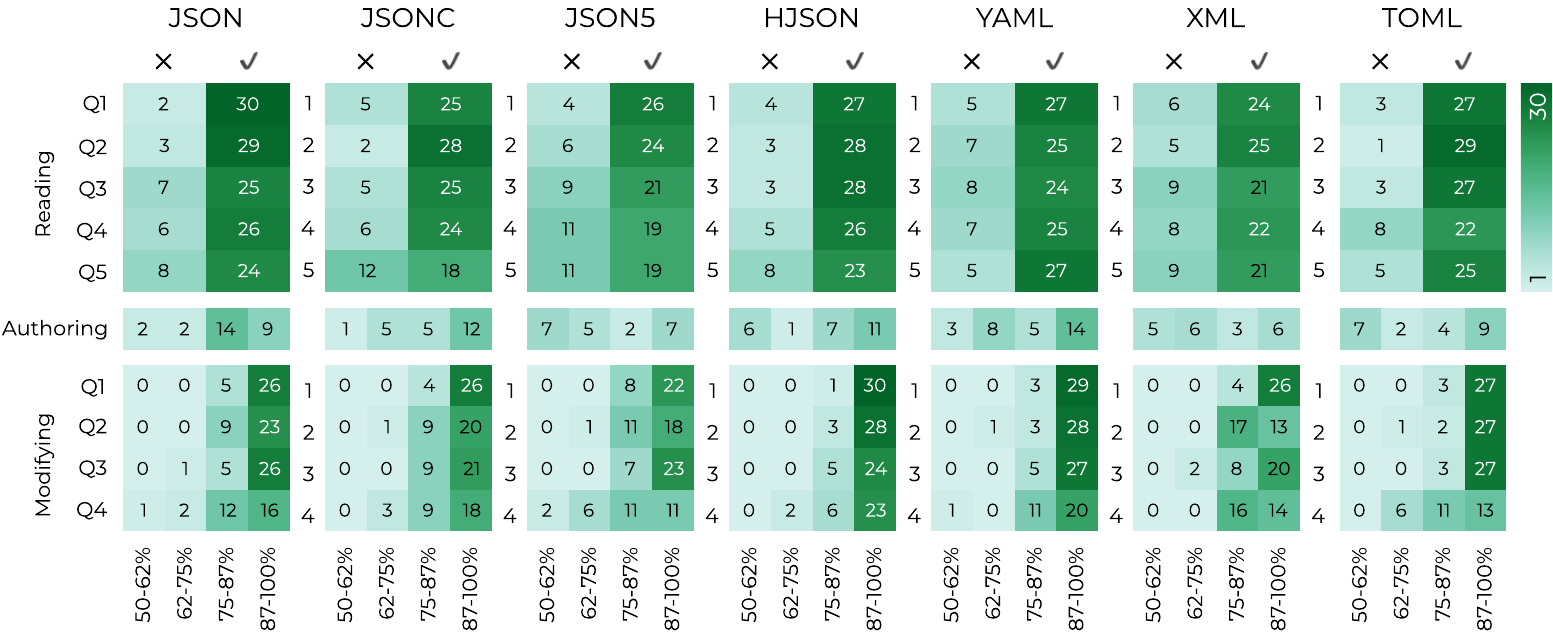}
    \caption{While there was variance in performance across formats, no one format stood out as best.
We note that \writing{} and \modifying{} had opportunities for partial credit, while \reading{} questions were pass or fail, which may account for the task type variance.
        Post-hoc pairwise comparisons showed statistically significant differences in the second \modifying{} subtask (field addition), in which TOML, YAML, and HJSON each outperformed XML ($p<0.01$).}
    \label{fig:task_trends}
    \vspace{-1em}
\end{figure*}

\begin{figure*}[t]
    \centering
    \includegraphics[width=\textwidth]{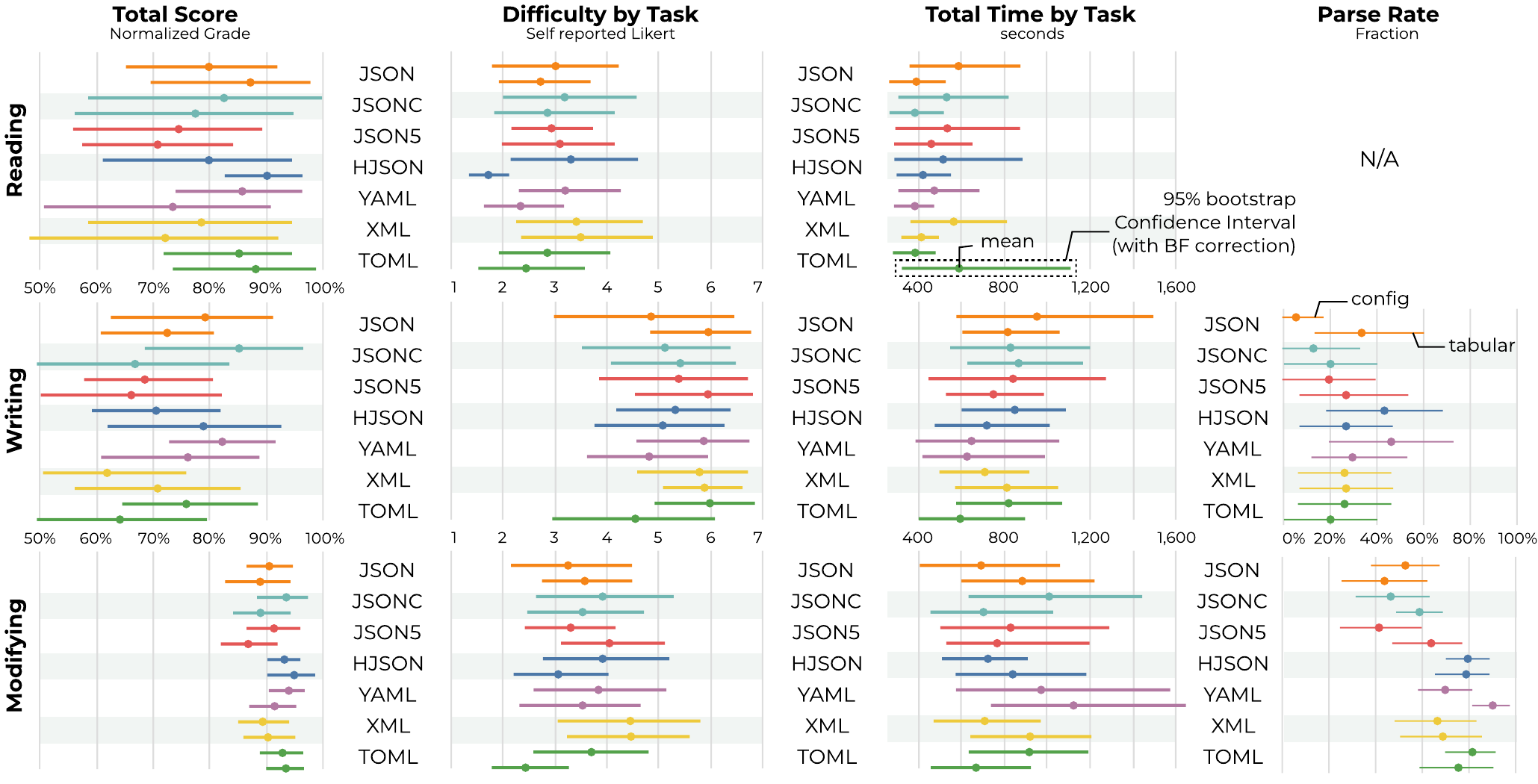}
    \caption{
        Across all tasks there is not a single dominant format, shown here by condition (format $\times$ task $\times$ context), each of which had $\geq$15 participants.
        Intervals are shown as bootstrapped 95\% confidence intervals with Bonferroni multiple comparisons correction.
        For \modifying{} and \authoring{} tasks, participants submitted freely edited text that either parsed or, more often, failed to parse, with parse rates varying across formats in ways that appear closely related to their grammar flexibility.
    }
    \label{fig:summary-results}
    \vspace{-1em}
\end{figure*}

\parahead{Limitations}
This study has several limitations. Given the constraints of our crowd-work study, tasks were necessarily simplified and might not capture all real-world scenarios (\eg{} debugging or learnability).
We drew from the general pool of participants on Prolific, rather than targeting expertise levels such as programming proficiency.
This sampling strategy was intentionally designed to minimize the confounding effects of pre-existing format preferences. While we did not see a substantial effect (\figref{fig:experience-heat}) of experience on our results, exploration of ability differences between novices and experts is worthwhile future work.
Inter-rater reliability of our grading scheme was established through calibration rather than independent parallel grading, future work could include formal IRR assessment.
Finally, manual grading may introduce bias. However, we applied consistent criteria across formats to ensure that any bias was systematic rather than favoring particular formats (\eg{} due to parser quality).

\subsection{Resampling JSONC \cfg{}}
\label{sec:Resampling}

During data validation using the autograder, we observed unexpectedly poor performance in the JSONC condition for the \cfg{} context. This was surprising given JSONC's minimal difference from JSON and the lack of similar effects in \tbl{} context. We examined potential procedural explanations (\eg{} AI use, skipped questions) but found no evidence to account for this deviation.
See appendix for details
To determine whether this reflected a systematic issue or an anomalous sample, we resampled the affected condition and got a more appropriate distribution.
We also manually graded in the question to ensure consistency, which replicated the anomalous pattern, suggesting the discrepancy was not due to grading mechanisms. While other factors cannot be fully ruled out, this points to sampling irregularity and so we report results using the resampled group throughout. This is akin to drawing a range of outliers and then excluding them. We include all data in our materials for transparency (see appendix).

\subsection{Results}

Next we describe our study's results, centered on performance differences by format and the effects of task.

\parahead{Performance differences are minimal by format}
\label{sec:perf}
We find no robust cross-task advantage for any format.
However, this null result requires some nuance: in specific \modifying{} tasks some formats (HJSON, YAML) outperformed alternatives.
We present these findings by task type, then examine potential explanations.
They are centered around a two-way ANOVA ($\alpha = .05$) with data contexts (\cfg{}, \tbl{}) and formats as fixed factors, testing main effects and their interaction on participant performance\footnote{We measured typing speed and considered normalizing task times relative to that rate.
    However, this adjustment did not alter the pattern of results and introduced additional complexity in interpretation. We report non-normalized typing times which more directly reflect task performance.}
and perceived difficulty.

First, we consider \reading{}. We observed no significant difference in performance, perceived difficulty, or task time across of the formats. Participants extracted information from structured data with comparable accuracy and speed regardless of notation---whether brace-delimited or indentation-based.

Next, we examine \authoring{}. Similarly, when participants translated an image to a structured textual representation, we found no measurable difference in performance. The high cognitive demands of this task appear to overwhelm any notational differences.

Finally, we look at \modifying{}{}. While this task predominantly held the same pattern, we observed one minor deviation. In particular, the \modifying{} subtask that required adding new fields (see \figref{fig:task_trends})---TOML, YAML, and HJSON yielded significantly higher performance scores than XML ($p<0.01$).
To further probe this finding, we examined parse success rates with the autograder in \writing{} tasks across formats, as it provides objective signals of syntactic flexibility to some extent. We observed that formats such as YAML, TOML, and HJSON tended to yield higher success rates compared to other more strict JSON families and XML, as shown in \figref{fig:summary-results}. These differences appear to align with the relative complexity and flexibility of format grammars. For instance, YAML and TOML support indentation-based structure and more flexible string representations, while HJSON relaxes requirements such as quotes and commas. Such affordances can reduce symbolic burden and mitigate editing errors under contextual support. In contrast, stricter formats (\eg{} JSON or XML), which impose rigid structural requirements, may increase the likelihood of syntax errors during writing.
Overall, the variances observed in the specific \modifying{} task suggest that formats which relax symbolic constraints and increase syntactic tolerance might effectively mitigate editing friction, providing a measurable advantage over stricter counterparts in specific interactive contexts.
While these trends do not manifest uniformly across all task types, they partially substantiate prevailing community discourse about syntactic strictness trade-offs.
To wit, the performance lag observed in XML (\figref{fig:task_trends}) resonates with common critiques of its verbosity and structural overhead.

However, we emphasize that these trade-offs do not consistently translate into measurable performance outcomes across the board. While specific formats exhibit statistically significant advantages under certain conditions, these benefits are not universal.
The benefits of syntactic permissiveness appear narrowly scoped to a small set of editing tasks where contextual cues support the modification process.

\parahead{Task Difficulty Overshadows Notation}
\label{sec:task}
Upon further investigation, we found that the influence of task difficulty overshadowed that of the notation itself, as the primary determinant of user performance. To elucidate these outcomes, we examined perceived difficulty across both data contexts and task types to characterize how cognitive demand governed the observed performance patterns rather than syntax alone.

First, we consider the difficulty related to the data context: \tbl{} versus \cfg{} as shown in \figref{fig:summary-results}.
We noticed that in \reading{} tasks, the \cfg{} was rated as more difficult and took longer to complete than the \tbl{} ($p=0.014$ for difficulty; $p=0.043$ for time), whereas format had no statistically measurable effect, suggesting that task context rather than format shaped perceived effort and completion time.
We conjecture that this difference stems from the relative unfamiliarity of the datasets: the \cfg{} dataset (\eg{} software configuration files) was likely less familiar to participants than the \tbl{} dataset (\eg{} student grades).
We suggest that most people have previously encountered \tbl{} data, whereas \cfg{} is comparatively rarer given participant backgrounds.
Specifically, 66\% of participants had no more than 2 years of programming experience (with 30\% having 0 years), and only about 17\% of participants rated themselves as ``very familiar'' with package.json files (which were used for \cfg{}).
Even with comparable accuracy, participants experienced \cfg{} tasks as more demanding and required additional time to complete.
This pattern suggests that format differences may only matter at a particular cross-section, where tasks are neither trivial enough to flatten performance nor demanding enough to overwhelm all participants regardless of notation. At the margins, format effects are attenuated.

Next, we examine the task-level difficulty, which follows a natural gradient of cognitive demand: \reading{} tasks generally require the least effort, \modifying{} tasks introduce moderate complexity, and \authoring{} tasks place the highest demands on recall, planning, and error correction. Per our results in ~\figref{fig:summary-results}, performance and perceived difficulty vary widely across tasks.
For the least demanding \reading{} and the most demanding \authoring{} tasks, no reliable score differences emerged across formats. This suggests that when tasks are either trivially easy or overwhelmingly difficult, format differences are unlikely to play a role.

To ensure comparability, we focus on  \authoring{} and \modifying{} tasks.
As shown in \figref{fig:task_trends}, performance diverges substantially across them.
Compared to \modifying{}, the \authoring{} task requires producing syntactically valid and structurally accurate content from scratch, placing higher production overhead and greater demands on recall, planning, and error correction---all of which likely contribute to the observed performance drop in \authoring{}. Within \modifying{}, however, we find that format differences are most apparent at intermediate levels of difficulty. At both extremes (the trivial modification of a single value and the high-complexity restructuring of data) no format outperformed the others.
This further underscores that format effects are attenuated at task difficulty extremes.

\begin{figure}[t]
    \centering
    \includegraphics[width=\linewidth]{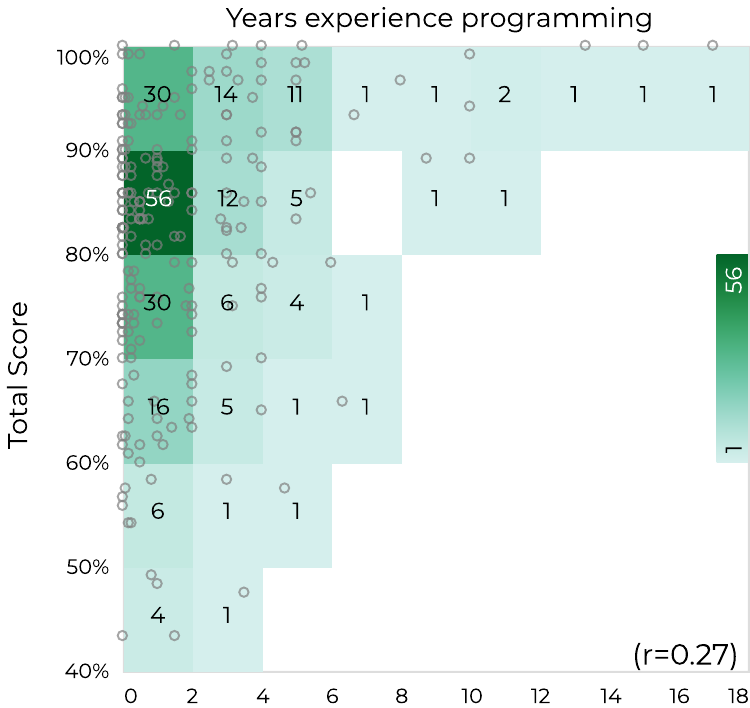}
    \caption{
        Self-reported experience distribution across participants. Although the pool skewed toward less experience, we observed only a weak relationship between experience level and task performance, suggesting that prior exposure did not confound format comparisons.
    }
    \label{fig:experience-heat}
    \vspace{-1em}
\end{figure}

To assess whether these results could be confounded by participants’ self-reported prior experience, we examined its effect on performance. Here we used a composite metric---total score---that sums scores from the \reading{} and \writing{} tasks to capture participants’ overall performance. As shown in ~\figref{fig:experience-heat}, we did not find any apparent influence on overall scores.
The limited sample size of highly experienced participants may reduce the sensitivity of this analysis and obscure more subtle effects. Further studies with a more balanced distribution of experience levels may be needed for better assessment.

These results indicate that no format confers a task-general advantage: if a single best \DSF{} existed, we would see it consistently across tasks.
Instead, any benefits are context-dependent. For instance, in our user study, format advantages emerge primarily in certain \modifying{} scenarios with permissive syntax and contextual cues, but fade at the extremes of difficulty---in \reading{}, \authoring{}, or the easiest and hardest \modifying{} subtasks. Overall, users' performance largely tracks task difficulty rather than notation. When tasks are trivial or highly demanding, outcome distributions compress and between-format differences are effectively attenuated.

\section{Interview Study}
\label{sec:rq2}

We now consider the question ``What factors drive usability other than notation design?'' via an interview study.

\subsection{Methodology}
We conducted semi-structured interviews with experienced developers to understand the usability issues of \DSFs{} in daily maintenance work, and their usage over time more generally.
We solicited participants on social platforms such as Twitter, Bluesky, and LinkedIn, filtering applicants to those with $\geq3$ years of relevant experience---yielding $N=\Ninterviews{}$ participants (1 using she/her and 8 using he/his).
Participants held a variety of roles, including full-stack development, front- or back-end development, and data engineering. All participants held at least a bachelor's degree in CS or a related field.
Each online interview lasted  $\sim30$ minutes.
Participants received a \$40 Amazon gift card as compensation.
Interviews were automatically transcribed and then reviewed by the first author for accuracy.
This study was marked exempt by the University of Utah IRB.
We complement these data sources by drawing on the comparisons and debates about  \DSFs{} that frequently appear on technically oriented platforms such as Hacker News, Reddit, and developer blogs. Echoing Barik \etal{}~\cite{barik2015hacker}, we view these sources as collections of practitioner perspectives and informal expert commentary.
We used affinity diagramming~\cite{hanington2012universal} of participant quotes to collaboratively cluster insights and surface recurring patterns in the interview data. The first author initially grouped excerpts, after which the project team met regularly to refine categories and resolve discrepancies.
More details can be found in the appendix.

\parahead{Limitations}
While the participant sample in our practitioner interview study included a range of professionals, it may not capture the full diversity of \DSF{} usage contexts.
Perspectives reported here may reflect specific organizational practices or individual experience, and should be interpreted as indicative rather than comprehensive. Coverage captured varied roles, domains, and cultural contexts could surface additional usability concerns not observed in this study.

\subsection{Results}

We find that conventions, preferences, and tool ecosystems are central to perception of usability---underscoring the influence of sociotechnical factors compared to notation alone.

\parahead{Conventions}
Standards or conventions held either locally (within teams or projects) or more globally (within developer ecosystems) significantly shape views about \DSF{} use and usability.
Rather than selecting formats based on intrinsic syntactic features, participants described deferring to what is commonly used or expected. This choice was often motivated by the availability of documentation, examples, and tooling support, making widely used formats easier to integrate. These observations echo Meyerovich and Rabkin's~\cite{meyerovich2012socio} findings, which emphasize ecosystem and community factors as central to adoption over purely technical characteristics.

\pdavid{} put it plainly, \interview{I just do whatever is the convention\ldots{}The best format is the one that follows convention---because the team can find more resources that way}. Adherence to familiar norms often takes precedence over critical evaluation of alternative formats.
Within teams, choices typically reflected established practice and the dominant technology stack. For instance, participants working in JavaScript environments usually defaulted to JSON. \pdominik{} noted that the widespread use of formats such as JSON and CSV stems less from their technical expressiveness than from their status as recognized defaults: \interview{Depending on who I work with, if the community agrees to use JSON or CSV, I will avoid using other formats like HJSON.}  Participants echoed this sentiment, noting that formats often gain adoption not due to technical superiority, but because they are widely supported and understood. To wit, \pyu{} mentioned: \interview{the primary consideration for choosing a format is that it must be in widespread use, and most programming languages should support it very well.}

Convention can sometimes override perceived technical flaws. Several participants acknowledged XML’s structural burden---\pyu{} described it as \interview{the most frustrating data format}, and \pallen{} noted: \interview{I try to avoid XML. It’s not popular, not easy to use, and to some degree, I think it’s outdated.} Yet its continued use is often shaped by existing conventions rooted in legacy systems and organizational dependencies. As \pyu{} admitted, \interview{We mainly use YAML for configurations, and occasionally XML---only because legacy systems require it.} Prior work similarly shows that long-standing infrastructure and institutional dependencies can sustain technologies beyond their technical prime~\cite{mahmoudi2024from}. A Reddit poster also echoed this sentiment, connecting XML's prominence with its history, saying \hn{Like it or not, XML has pretty much gotten to be a universal data exchange format.}~\cite{vplatt}
These accounts suggest that convention is intertwined with the tooling and infrastructure that accumulate around widely adopted formats. As formats become entrenched, they attract continued investment in libraries and integration support, making them harder to replace. This co-evolutionary dynamic makes convention and tool ecosystem mutually reinforce each other, stabilizing formats despite notational drawbacks.

Long-term adoption depends less on notational dimensions than on alignment with shared expectations, community norms and tooling ecosystems. As one Reddit poster remarked on what ultimately makes a format the best option: \hn{The real-world answer is also: what are things already using.}~\cite{AnimaLepton} A similar sentiment appears in the HJSON repository, where a contributor noted: \hn{The main way to get the format noticed is in real-world usage. It would only take one major project to use it for it to explode in popularity.}~\cite{Joyless}
Once a format becomes a default, ensuing pressures to follow convention and ensure compatibility can easily outweigh the notational usability advantages of competing designs.

The importance of convention does not mean negating the role of syntax features. Certain formats align more naturally with particular data structures due to their inherent characteristics, which can in turn contribute to widely shared conventions. For example, CSV's is well-suited for flat, two-dimensional datasets but ill-suited for hierarchical data; as \pstun{} observed, \interview{It makes complex data representation difficult.} \pyu{} echoed this view: \interview{If I'm working with, like, two-dimensional data, I will use CSV, even if I'm working in a JavaScript project.} Such concrete feature alignment was frequently cited among our participants as a reason for favoring one format over another in specific contexts. Besides, the ability to convey information beyond strict syntax (\eg{} through comments) can influence format adoption when teams need to embed human-readable explanations.
For instance, \pdaniel{} remarked JSONC could be useful: \interview{If you want to leave some comments for other developers}, and a Reddit poster similarly noted: \hn{there's no way to put comments in JSON, which, IMO, are crucial if you're using it to store any kind of user-editable configuration data.}~\cite{minneyar}
However, the absence of such support in strict formats like JSON has not prevented its widespread use. Instead, it has generated a host of workarounds~\cite{jsonComments}, including dedicated ``comment'' fields, separate documentation files, or even repeated field names that exploit JSON's uniqueness guarantee. These practices further illustrate that although syntax can shape conventions by enabling certain capabilities, entrenched conventions can prove stronger than notational limitations, driving practitioners to adapt around shortcomings rather than adopt alternative formats.

\parahead{Preferences}
\label{sec:prefs}
We find that preferences shape perceptions of \DSFs{}, where preferences often center on aesthetic judgments of readability and structure.
For example, the trade-offs of \DSFs{} between indentation-based structures and explicit delimiters in shaping readability, which users evaluate through their own aesthetic sensibilities.
In community discussions, some users argue that YAML's indentation makes structure more visually apparent and thus easier to read than JSON's use of braces---valuing one form of notational visibility. As one Reddit poster summarized this idea noting that  \hn{Proper indentation makes code blocks stand out and thus easier to read.}~\cite{desrtfx}
Another Hacker News poster observed that \hn{Personally I mostly use indent to read code, so requiring that the indent matches the semantic nesting makes it much easier for me to understand.}~\cite{kevincox}
This suggests that indentation-based formats like YAML may strengthen the connection between visual layout and hierarchical structure, possibly by reducing visual clutter and symbolic overhead.
We found it notable that these perspectives were rooted in personal opinion about readability, and not something more oncrete such as linguistic background or reading disabilities (\eg{} dyslexia).

Yet, this preference for indentation is not universal as indentation can also potentially yield errors via misaligned or inconsistently applied indentation. Others strongly prefer explicit delimiters, as in JSON's curly and square braces: \hn{What is the obsession with removing braces? I will never find the lack of clear demarcations (relying on indent) easier than braces.}~\cite{diob} \pdavid{} offered a similar perspective \interview{I prefer JSON, because I don't really like indented formats like YAML\ldots{}and I prefer just using curly braces, because it's more clear.} On the other hand, formats with explicit delimiters, like JSON, make structural boundaries apparent often at the cost of a larger concrete syntax. For JSON this leads to errors related to forgotten delimiters (\eg{} commas) as people usually complain: \hn{This missing comma ruined my day.}~\cite{rvscode}

Users vary in this trade-off---some users find explicit markers more straightforward, while others value the simplicity and spatial economy of whitespace-based structures.
These divergent views highlight that preferences toward different notational trade-offs are subjective, and do not imply the inherent superiority of any single design choice---as our quantitative results suggest that their impact on common tasks' performance is minimal.
These preferences may be entwined with programming language familiarity, as a Reddit poster put it: \hn{I don't like all the curly braces and [explicative]. I hate hate hate JavaScript and anything closely related to it.}~\cite{anantnrg}
Regardless of the origins of those subjective inclinations, they seem to be influenced by the task context and the scale of data.
Format aesthetic benefits tend to fade as task complexity increases.
For instance, one practitioner noted
\hn{YAML files are easier to change by hand. As the configuration files are currently mostly edited by people.}~\cite{gr2m}
Similarly, \prahat{} commented that he is satisfied with TOML and YAML for routine edits as he can \interview{change the value safely} whereas JSON often forced him to wrestle with braces or auto-formatting quirks.
However, these perceived usability benefits might not be extended to more complex contexts.
As one Hacker News poster complained: \hn{I hate YAML with a passion...the whitespacing and awkward syntax is really just too frightfully painful on large configs.}~\cite{vorpalhex} Another practitioner also noted this point: \hn{indentation to be confusing.... especially if there are multiple levels or longer function blocks.}~\cite{ErikAronesty}
This pattern echoes our quantitative finding that format differences attenuate at difficulty extremes.

\parahead{Tooling and Ecosystem}
\label{sec:tooling}
Beyond conventions or individual preferences, the surrounding tool ecosystem, such as linters and IDE integration, plays a decisive role in shaping \DSF{} use. As \pdaniel{} remarked: \interview{JSON tends to be user-friendly, not because of the format itself, but because there are so many tools to work with it},
suggesting differences in ecosystem support often determine which formats prevail in practice.

Participants frequently described relying on formatters, linters, and automated generation tools to mitigate notational fragility. As \pdominik{} explained, while formats such as YAML and JSON each present syntactic challenges these issues are often mediated by tooling; he noted that he \interview{often write JSON not by hand, but it's programmatically generated.}
In such contexts, tooling reduces the practical salience of syntactic differences, explaining why multiple formats with distinct surface forms coexist and remain viable in practice.
Early error detection was seen as especially critical for routine work, where mistakes can silently propagate. \interview{I've made many mistakes in configuration files.}, As \pdominik{} put it: \interview{The formats themselves are fine---the real need is for better tooling, like libraries that surface errors and tools that show problems as early as possible.} Editor tools such as Prettier, ESLint, and VSCode were credited with improving confidence and standardizing practices, as \pallen{} noted: \interview{We frequently use Prettier and ESLint together in front-end development\ldots{}ESLint detects style issues.} Conversely, the absence of robust tooling created fragility in day-to-day use, especially for whitespace-sensitive formats like YAML. \prahat{} recalled: \interview{When copying code from StackOverflow or ChatGPT, space and tab issues can prevent the code from working.}
\pdavid{} explains that the lack of debugging support made YAML feel like \interview{a horrible platform for configuring.}
Compounding this issue, heterogeneity of different platforms can create unpredictable environments where a file might be validated by one tool may be rejected by another.
As one developer noted: \hn{Depending on my editor (visual code or vim) or my linter (yamllint python package or this website), I end up with a valid or invalid yaml file.}~\cite{Jonas}
Sometimes cryptic error messages (\eg{} ``unexpected token'')  further hinder debugging.
One example of tool-based frustration comes from so-called round-tripping, where saving or reformatting files can reorder content or strip annotations. As one poster recalled: \hn{If an automations.yaml file contains comments, writing to it with Automation Editor purges all the comments.}~\cite{123} These frustrations shape perceptions of formats.

The impact of limited tooling was also evident in our crowd-work study.
The post-task NASA-TLX survey yielded a relatively high average workload score of $67 \pm 15$ (out of 100). Participants reported unexpectedly high physical demand ($66 \pm 31$) for a primarily typing-based task. Post-study comments noted that our policy prohibiting copy and paste (per Prolific's~\cite{prolific_llm_guidance} recommendations for discouraging AI-usage) made the task feel unnecessarily difficult.
One participant explained \interview{I have some experience in writing code SQL/Python and I was still questioning myself. It takes a while to be comfortable with syntax and honestly, most coders copy and paste a lot}.
Highlighting how interface-level aids offset the higher friction aspects of \DSFs{}.
We also found that participants who used aids such as search (\eg{} \texttt{Cmd+F}) seemed to consistently achieve higher task performance.

Considering our earlier finding that format differences had minimal influence on outcomes, this pattern aligns with prior work in programming languages showing weak relationships between intrinsic language features and real-world performance. For example, Berger \etal{}~\cite{berger2019impact} observed that language features such as static typing or paradigm (\eg{} functional versus object-oriented programming) show only weak correlations with software defect rates, suggesting that intrinsic notational properties are poor predictors of real-world performance, while Meyerovich and Rabkin~\cite{meyerovich2012socio,meyerovich2013empirical} show that developers rarely select languages based on specific features alone.
Further this suggests that usability depends less on notational design than on how participants engage with the available tool ecosystem.

\section{Discussion}
\label{sec:discussion}

This work examined whether notational differences meaningfully shape \DSF{} usability across common tasks, and what factors beyond syntax may influence experience.
We find no evidence of a single ``best'' format from a notational perspective. Instead, usability of \DSFs{} is governed by a complex interplay of sociotechnical factors.
The results of our crowd-work user study suggest the notational properties alone have a modest impact on usability, and explain why community debates often remain polarized---the lack of consistent empirical differences invites developers to fill the evidentiary gap with aesthetic preferences and leaves room for conflicting interpretations.
Our interview study
suggests that the usability of \DSFs{} should be understood within broader organizational and infrastructural contexts beyond notation alone.
Taken together, future work could consider more realistic, more full-featured editing environments such as those with linters, formatters, and IDE support. Comparing those results with the present study would help distinguish intrinsic notational effects from the tooling and ecosystem factors that shape \DSF{} use in practice.

Our findings also align with the enduring debate about the feasibility of a universally optimal language. Rather than a linear progression toward a single ``best'' paradigm, Stefik \etal{}~\cite{stefik2014programming} characterize the field as oscillating between specialized solutions to contexts. This echoes our observation that task context shapes how users perceive and evaluate \DSFs{}.
Meyerovich and Rabkin~\cite{meyerovich2012socio, meyerovich2013empirical} find that language success is mediated by sociotechnical factors---such as tooling, libraries, and community---rather than by syntactic features.
Our results reinforce this pattern: ecosystem support and conventions shape perceived usability more than notational design, whose influence appears more limited than commonly assumed.
This resonates with Norman’s concept of ``knowledge in the world''~\cite{norman2013design}, where usability stems not only from internal design, but from how well the artifact fits into its broader environment. Formats like JSON or CSV were favored less for expressiveness than for the incentives surrounding them.

Collectively, these insights reinforce the importance of understanding language-like systems within broader sociotechnical environments rather than static technical choices.
The usability of a \DSF{} emerges through its integration with tooling ecosystems, where infrastructure can buffer syntactic limitations and shape practical experience.  As one developer evocatively remarked: \hn{I'll still write lots of YAML, and JSON, and all that other stuff. They all have strengths and flaws. Such is life.}~\cite{brews}
Multiple formats remain viable not because their notational quirks or functional gaps are resolved, but because tooling and conventions render them manageable in practice.
In practice,  designers should treat syntax as one component of a broader system rather than the primary lever of improvement---echoing Jakubovic \etals{}~\cite{jakubovic2023technical} broadening of the focus of the Cognitive Dimensions of Notation~\cite{green1989cognitive} to encompass whole systems. Investments in enhancing linting tools, formatters, IDE support, and migration tooling may yield greater usability gains than introducing new syntactic variants.

Format adoption follows prevailing conventions rather than isolated technical merit. Once established, feedback loops of tooling, documentation, and onboarding stabilize a format's position—making ecosystem inertia and migration costs as important as notational design. Supporting migration pathways and shared tooling may more effectively facilitate adoption than syntactic refinement.

The studies in this work focused on western users (the crowd work study) and professionals (the interview study).
Echoing Hermans \etal{}~\cite{hermans2024case}, future work should examine who \DSF{} designs benefit and center.  Who do they center as the idealized user?
For instance, does notational readability vary across linguistic and cultural contexts? Do right-to-left and left-to-right language communities engage with (or perceive) notations differently? While our results suggest that contextual and infrastructural factors dominate, the weight of these cultural factors is an open question.

Future work should investigate domains where multiple, functionally equivalent notations coexist. For example, understanding the usability of the various flavors of mark up languages (\eg{} Markdown, reStructuredText, AsciiDoc) is worth investigating---though our results suggest such effects may be limited.
Beyond these simple formats, systems involving non-identical but similar semantics are also of interest---such as the difference in visualization notations~\cite{kruchten2023metrics} or web development frameworks.
Evaluation of the usability of s-expressions may also be useful, as they are often decried as a poor format~\cite{wheeler_readable_problem}. Rhombus~\cite{flatt2023rhombus} offers an interesting natural experiment to explore usability and adoption effects both without (Rhombus) and with (Racket) parentheses---although adoption, as we stressed throughout, is deeply linked to sociotechnical dimensions and analysis will be non-trivial.
A notable recent strain in \DSF{} development has been developing formats that explicitly support the limitation of LLMs, as exemplified by TOON~\cite{toon} (Token-Oriented Object Notation) and LEAN~\cite{lean-format} (LLM-Efficient Adaptive Notation).
This raises a new design tension: balancing token-efficiency against human readability.
Our results suggest that the impact on human dimensions will be minimal---however, we note that TOON and LEAN take a greater abstraction leap (roughly combining YAML and CSV) than the modest syntactic differences considered.

While these findings are unlikely to settle every debate about \DSFs{}, as such issues are primarily driven by (community) preference, they signal a need to reframe conversations away from syntax-centric comparison and toward evaluation of the sociotechnical ecosystems in which formats operate.

\section*{Acknowledgments}
We thank our participants for sharing their time and insights with us.
We thank the ReVISit team for their support in making this work possible, as well as everyone who piloted this study including Devin Lange, Ben Greenman, Elham Ghelichkhan, and Whanhee Cho.
This work was supported in part by NSF award \#2402719.

\bibliographystyle{IEEEtran}
\bibliography{reference}



\onecolumn{}

\appendix{}

\section{Appendix}

\begin{figure}[t]
    \centering
    \includegraphics[width=\textwidth]{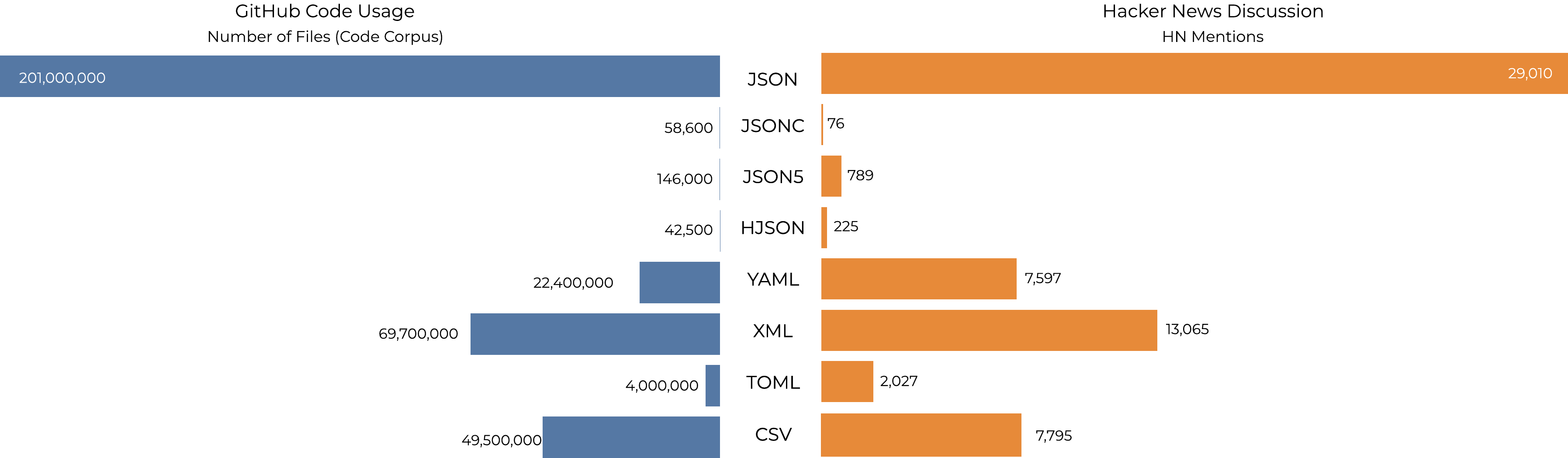}
    \caption{We assessed format popularity using two sources. For community engagement, we used the Hacker News Search API to retrieve the top 500 stories containing format-related keywords (JSON, JSONC, JSON5, HJSON, YAML, TOML, XML, CSV) along with their nested comments, and counted occurrences of those keywords. SON dominates both metrics, while newer variants (HJSON, JSON5) remain niche despite community interest. For prevalence in practice, we queried GitHub repositories and measured the frequency of files with corresponding extensions.}
    \label{fig:format_choice}
\end{figure}

\begin{figure}
    \centering
    \includegraphics[width=\linewidth]{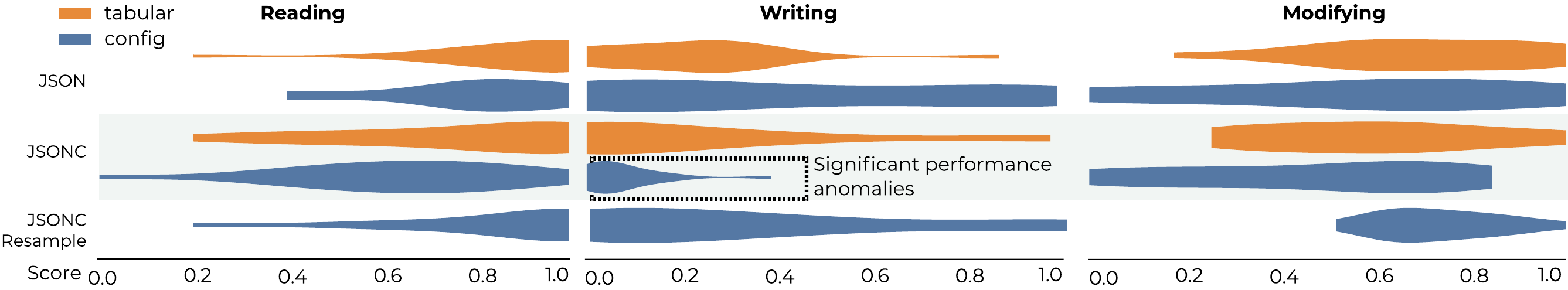}
    \caption{In the initial run of our crowd work study, we found an extreme performance difference between JSON and JSONC for \cfg{}---based on our initial automated grading scheme. This performance disagreed with both the performance on JSON (which hypothetically should be only minimally affected by the addition of comments) as well as the results for \tbl{}.
        To examine whether or not this result was anomalous, we resampled that condition, yielding results that more closely agreed with the rest of the set. We use the resampled values in a manner analogous to pruning an outlier.
}
    \label{fig:json-c-confusion}
\end{figure}

\begin{figure}[t]
    \centering
\includegraphics[width=0.3\linewidth]{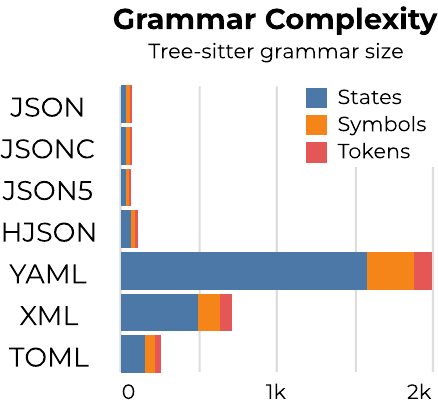}
    \caption{Format flexibility is approximated by the size of each Tree-sitter grammar, measured here by its numbers of states, symbols, and tokens. Although grammar size depends partly on implementation choices and could potentially be optimized, the large relative differences provide a useful proxy for the syntactic complexity that a parser must represent. The YAML grammar corresponds to YAML 1.2. However, the study used only features shared by YAML 1.1 and 1.2, so this version difference did not affect task evaluation or grading.}
    \label{fig:complexity}
    \vspace{-1em}
\end{figure}

\begin{figure*}[t]
    \centering
\includegraphics[width=\linewidth]{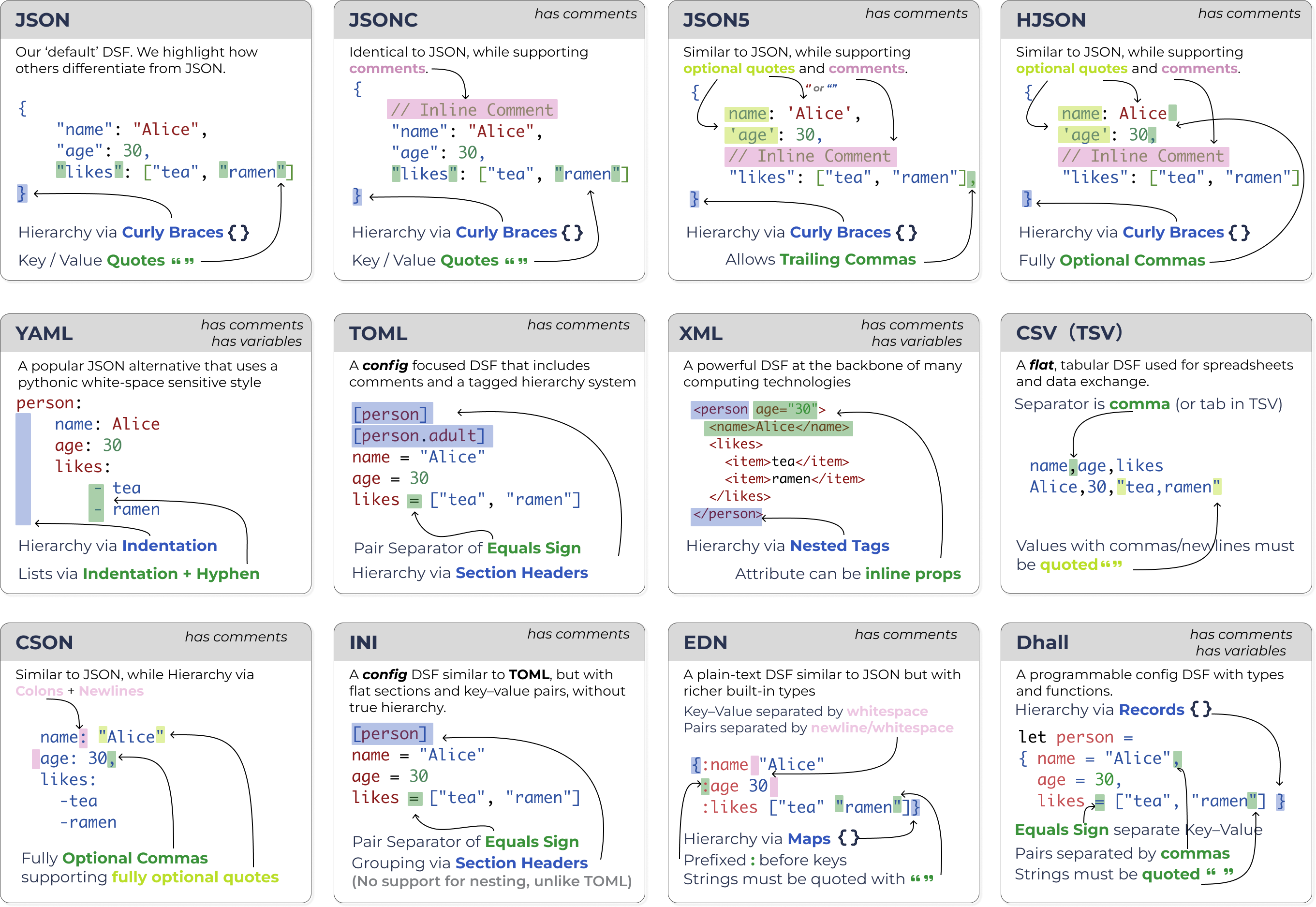}
    \caption{The studies in this work, JSON, JSONC, JSON5, HJSON, YAML, TOML, and XML---as well as CSVs where appropriate. Beyond these basic syntactic features, YAML and XML include a variety of advanced features, such as variable-like abstractions, which are often unused or ignored.
While CSV was part of our broader analysis, it was not used in the crowd work user study. Other formats were excluded from the main analysis due to overlapping functionality (\eg{} TSV vs. CSV), limited contemporary adoption (\eg{} INI), or relatively low visibility and discussion in developer and open-source communities (\eg{} EDN, Dhall, and CSON).}
    \label{fig:baseball-cards}
\vspace{-1em}
\end{figure*}

This appendix includes a variety of additional information that is relevant to this work, but was not included in the main body of this work.
In \autoref{fig:format_choice} we review the popularity of the \DSFs{} of interest on HackerNews and GitHub.
In \autoref{fig:json-c-confusion} we highlight the performance anomaly that led us to resample one of the conditions.
In \autoref{fig:complexity}, we compare their Tree-sitter grammar sizes as a proxy for the syntactic complexity represented by each parser.
In \autoref{fig:baseball-cards} we list all considered formats, including those excluded from the study due to practical constraints.
\autoref{tab:thematic_analysis} shows the results of our thematic analysis.
In \autoref{tab:grading_rubric} we describe the grading rubric we used to evaluate the \writing{} tasks in our crowd work user study.
In \secref{appendix:cdn}, we provide a replication of the cognitive dimensions analysis based on social media comments data as supplementary material.
In \secref{appendix:interview-guide} we include our interview instrument from our semi-structured interview study.
An example of our crowd work study can be found at \ourhref{https://dsf-study.netlify.app/a-usability-study-config/}{here}.
Additional materials, such as the code for our crowd work study can be found at  \osf{}.


\begin{figure*}
    \centering
    \includegraphics[width=\linewidth]{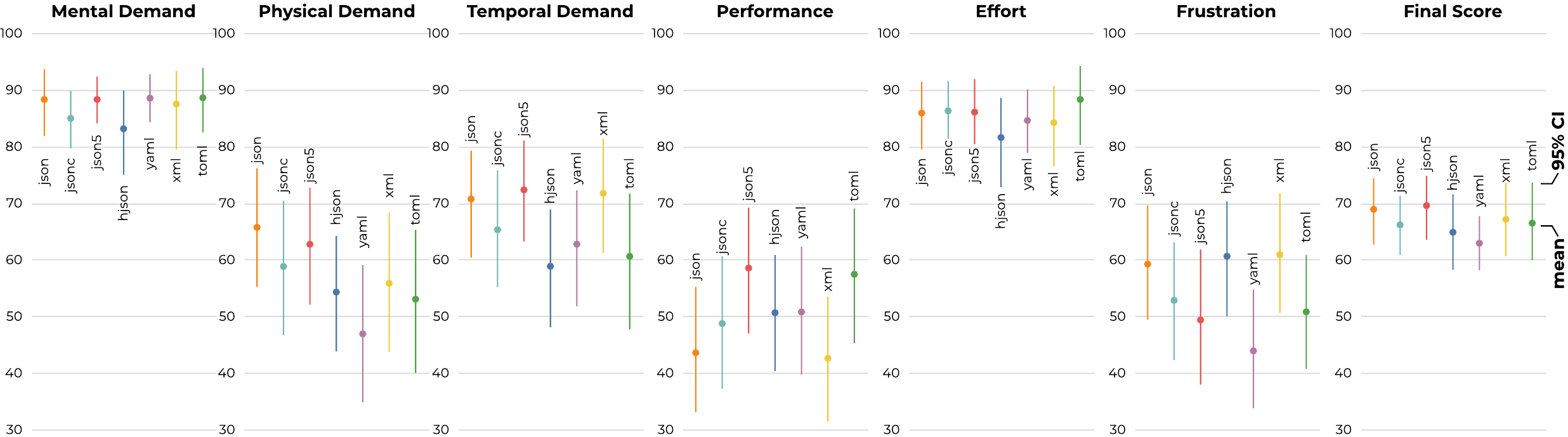}
    \caption{After the completion of the main part of the crowd work study, participants filled out a NASA TLX. The average across all formats was $67\pm15$, suggesting a high perceived load. Of particular note is the exceptionally high physical load, which is surprising for a typing based activity. It is possible that this may be related to our disallowing copy-paste (to reduce AI usage). }
    \label{fig:tlx}
\end{figure*}

\begin{figure}[!htbp]
    \centering
    \small{}
    \begin{tabular}{|p{1.5cm}|p{1.7cm}|p{13.4cm}|}
        \hline
        \textbf{Theme} & \textbf{Sub-theme}                                      & \textbf{Example Quotes} \\
        \hline
        \multirow{3}{*}{Conventions}
                       & norms within teams
                       & \begin{minipage}[t]{\hsize}
                             \begin{itemize}
                \item \pyu{}: \interview{We mainly use YAML for configurations, and occasionally XML---only because legacy systems require it.}
                \item \pdominik{} : \interview{Depending on who I work with, if the community agrees to use JSON or CSV, I will avoid using other formats like HJSON.}
            \end{itemize}
                         \end{minipage}                                                      \\
        \cline{2-3}
                       & norms within ecosystems
                       & \begin{minipage}[t]{\hsize}
                             \begin{itemize}
                \item \pyu{}: \interview{the primary consideration for choosing a format is that it must be in widespread use, and most programming languages should support it very well.}
                \item \pdavid{}: \interview{I just do whatever is the convention\ldots{}The best format is the one that follows convention---because the team can find more resources that way}
            \end{itemize}
                         \end{minipage}                                                      \\
        \cline{2-3}
                       & notation \newline properties
                       & \begin{minipage}[t]{\hsize}
                             \begin{itemize}
                \item \pstun{}: \interview{CSV makes complex data representation difficult.}
                \item \pyu{}: \interview{If I'm working with two-dimensional data, I will use CSV, even if I'm working in a JavaScript project.}
                \item  \pdaniel{}: \interview{JSONC could be useful if you want to leave some comments for other developers}
            \end{itemize}
                         \end{minipage}                                                      \\
        \hline

        \multirow{3}{2cm}{Tooling \&\newline Ecosystem}
                       & Error \newline Detection and \newline Debugging Support
                       & \begin{minipage}[t]{\hsize}
                             \begin{itemize}
                \item \pdominik{}: \interview{The formats themselves are fine---the real need is for better tooling, like libraries that surface errors and tools that show problems as early as possible.}
                \item \pallen{}: \interview{Prettier is very frequently used in front-end linter; actually, we also use ESLint, it was used to detect some basic errors in style information. We always use them as combination. Also, ChatGPT is another very good tool now. I think some AI tools are already merged, merging into some IDEs, like VSCode, If we want some AI tools to help us do some data serialization, I think it's a good choice.}
                \item \pdavid{}: \interview{ I think YAML is a horrible platform for configuration. I think they give too much power to the YAML file that it like I said, it becomes a programming language, and then it's hard to diagnose when something goes wrong, because it doesn't have all the support or debugging support that our actual programming language has, I think they should just make make us write like a script, instead of configuring with yaml.}
                \item \prahat{}: \interview{I like YAML a lot because I use it most of the time. But again, for team collaboration, one of the things I shared with you is the space versus tab, and it happens most of the time when I copy any code from Stack Overflow or any other place, even from ChatGPT, and it doesn't work.}
            \end{itemize}
                         \end{minipage}                                                      \\
        \cline{2-3}
                       & Workflow \newline Integration
                       & \begin{minipage}[t]{\hsize}
                             \begin{itemize}
                \item  \pyu{}: \interview{I think JSON actually came from JavaScript. So whenever I use JavaScript, I prefer to work with it because it has native support.}
                \item \pdavid{}: \interview{ I think one of the main benefits of JSON is that almost every language has a JSON parser. Yeah, that's, that's like, one of the main reasons why you would use it for like, code-to-code communication.}
                \item \prahat{}: \interview{ So I kind of like working with C++ and Python. For both of them, I find fantastic support from both the community and the support of software available, the most expensive ones are for JSON. That's why I use JSON, and also for CSV. }
                \item \pmorris{} \interview{This kind of spreadsheet structure(CSV), which always makes it easy to edit, like in tools like Excel, Google Sheets, or text editors, is mostly light and written fast.}
                \item \pdaniel{}: \interview{JSON tends to be user-friendly, not because of the format itself, but because there are so many tools to work with it.}
                \item \pdominik{}: \interview{In CSV, coding is hard. In YAML, the indentation is hard. In JSON, you know, you have to put the comment below some quotes and so on. I often write JSON, not by hand, but it's programmatically generated, and so whatever. But then you want to have maybe a consistent format, so I try to use different formatters.}
            \end{itemize}
                         \end{minipage}                                                     \\
        \hline

        \multirow{3}{2cm}{Individual\newline{}Preferences}
                       & Indentation \newline vs. \newline Delimiters
                       & \begin{minipage}[t]{\hsize}
                             \begin{itemize}
                \item \pdavid{}: \interview{But personally, I prefer JSON, because I don't really like indented formats, like YAML, you have to use indentation to show the difference between the hierarchy or the code blocks, and I prefer just using curly braces, because it's more clear.}
                \item \pdaniel{}: \interview{As someone who works with and generally likes infrastructure in the cloud, it's more than I work with YAML as configuration by format, and the significant white space is kind of cool, because it means that you can do it without adding quoters, curly braces, and all that sort of stuff. You can have objects and lists and that sort of stuff, which is kind of nice, but especially when templating YAML for Kubernetes, and then, just like having an extra space somewhere, the whole thing baking is so annoying. So I have a love and hate relationship.}
                \item \pmorris{}: \interview{For me, it's(YAML) simple and easier than JSON, because it mostly avoids brackets and uses indentation.}
            \end{itemize}
                         \end{minipage}                                                     \\
        \hline
    \end{tabular}
    \caption{Themes resulting from our analysis of our semi-structured interview study. We employed affinity diagramming~\cite{hanington2012universal} as our primary method to synthesize interview data. The first author began with an open coding pass, extracting salient excerpts from transcripts and recording those quotes as cards in Figma, and clustering inductively by semantic similarity. For example, grouping reflections on conventions or frustrations with tool support. The research team then met regularly to review and refine the clusters, re-labeling or merging groups as needed and resolving discrepancies through consensus. After several iterations, we concluded three major themes: Conventions, Preferences, and Tooling and Ecosystem. These themes capture recurring participant concerns, with several sub-themes identified within each major category to further organize the findings.}
    \label{tab:thematic_analysis}
\end{figure}

\begin{figure}[t]
    \centering
    \begin{tabular}{p{0.15\linewidth} p{0.45\linewidth} p{0.3\linewidth}}
        \toprule
        \textbf{Dimension} & \textbf{Error Type}          & \textbf{Deduction}              \\
        \midrule
        Syntax             & Missing symbol               & -1 if occurs                    \\
                           & Extra symbol                 & -1 if  occurs                   \\
                           & Misused symbol               & -1 if  occurs                   \\
                           & Misplaced symbol             & -1 if  occurs                   \\
                           & Overall structure confusing  & -1 discretionary                \\
        \midrule
        Semantics          & Missing or incorrect content & -1 if occurs                    \\
                           & Extra or irrelevant content  & -1 if occurs                    \\
                           & Serious missing content      & -1 if occurs (greater severity) \\
                           & Serious extra content        & -1 if occurs (greater severity) \\
                           & Meaning severely unclear     & -1 discretionary                \\
        \bottomrule
    \end{tabular}
    \caption{Grading rubric used to evaluate responses of \writing{} tasks in the crowd work study. Each dimension is scored on a 0–5 scale after deductions.}
    \label{tab:grading_rubric}
\end{figure}

\subsection{Social Media Analysis}
\label{sec:social-media-meth}

Comparisons and debates about different \DSFs{} frequently appear on technically oriented platforms such as Hacker News, Reddit, and developer blogs. Echoing the approach of Barik \etals{}~\cite{barik2015hacker}, we view these sources as collections of practitioner perspectives and informal expert commentary. Prior work has similarly drawn on such discourse to examine attitudes toward programming-related technologies~\cite{sarkar2022like,sarkar2022end}.
To broaden our understanding beyond our own experiences and the structured settings of our empirical studies, we conducted a lightweight qualitative review of online discussions concerning \DSFs{}.

Using the Hacker News API, we retrieved 2,966 posts and comments mentioning formats of interest (\eg{} “JSON,” “YAML,” “XML,” “TOML”). We complemented this dataset with Reddit threads and developer blog posts identified through keyword searches (\eg{} “JSON usability,” “config files,” “tabular format”) and snowball sampling. We excluded off-topic material where formats were mentioned only tangentially. Selected excerpts were manually reviewed and thematically organized by the first author, with periodic discussion among the research team.

These discussions reveal that perceptions of usability often arise from trade-offs rather than single design features. For example, YAML’s indentation-based structure is frequently described as making hierarchical data easier to read, yet also as increasing the risk of subtle formatting errors. Similarly, some developers praise JSON5 for features such as trailing commas and comments, while others argue that these additions introduce ambiguity or reduce clarity. Such exchanges illustrate that community judgments about usability are shaped by context, familiarity, and competing practical priorities, rather than by any single syntactic characteristic.

\section{Evaluation of Data Serialization Formats Across Cognitive Dimensions of Notations}
\label{appendix:cdn}

Cognitive Dimensions of Notations ~\cite{green1989cognitive} are a set of dimensions for analyzing the notations embedded in a system.
Roughly, notations are the aspects of a system which the user can use to compose new functionality. In contexts like \DSFs{} the syntax is essentially synonymous with notation, although in other domains, such as visual programming languages the connection is less direct.
We adopt this framework as our analytical foundation throughout our work to help us better understand the notational variances of \DSFs{}.
Some dimensions in the CDN framework---such as \cdn{Abstraction Gradient} and \cdn{Diffuseness}---can be relatively direct and intuitive when compared across the specifications of different \DSFs{}.
However, others---such as \cdn{Error-Proneness} and \cdn{Hard Mental Operations}---seem to be highly context-dependent and often shaped by users' subjective experiences.
To navigate this complexity, we draw on comments from large-scale developer discussions on online platforms (\hn{quoted like so}).
These include instances and anecdotes from practitioners experiences that complement our heuristic analysis, helping to contextualize and substantiate how different \DSFs{} perform along dimensions that are otherwise difficult to assess abstractly.
We now describe our analysis of each dimension, drawing prior definitions~\cite{cdn-wiki} of the dimensions.

\parahead{\cdn{Abstraction Gradient}} \emph{What are the minimum and maximum levels of abstraction exposed by the notation? Can details be encapsulated?}

\DSFs{} have very limited collections of abstractions as their remit is primarily focused on the storage of data.
We identify three core abstractions present in this domain: hierarchical data (an abstraction of tabular data), lightweight variables, and schemas.

First, consider hierarchy.
While all data could conceptually be represented as arrays of arrays, a common convention in \DSFs{} is to include key-value records which typically enforce uniqueness among keys at parse time, as means to simplify interacting with the data.
Among the formats we examine in our studies, all except for CSV include such a hierarchical data representation.
As a result, CSV is generally not considered for representing nested or structured data---although it can be improvised, such as by including serialized JSON a values, although this highlights the limitations of this notation.
This point is well known: one Stack Overflow user observed that \hn{CSV is made to store data as a two-dimensional array ... I don't know how you would format your CSV to create nested arrays ... the CSV format is simply not suitable for that.}~\cite{user8849929}

Next are variables.
Among the formats we examined, only XML and YAML mechanisms for reusing structures as part of its core syntax.
YAML's notion use of anchors and aliases enables the definition of shared configurations that can be referenced throughout a document---an affordance particularly valuable in contexts where reducing duplication and enhancing maintainability are critical. As one Stack Overflow user observed:
\hn{One notable benefit is to avoid duplication by defining a reusable block of data (Anchor) and referencing that block elsewhere (Aliases) in the YAML file.}~\cite{ColinWa}
Similarly, XML has a notion of entities which enable specification of values and literals.

Lastly are schemas.
Schemas exist for a variety of the \DSFs{} examined, although primarily this support arises from JSON Schema (for JSON and its variants) or XML Schema (for XML).
These schemas introduce a type-style validation or contract checking, that enables validation of the form of a \DSF{} file.
For instance, in enterprise environments, developers often rely on these schemas to catch upstream issues or validate client-server interactions.
In addition, schemas can be used as forms of documentation (such as for type hints) as well as the basis for constructing user interfaces~\cite{mcNutt2023editors}.

\parahead{\cdn{Hidden Dependencies}} \emph{Are dependencies between entities in the notation visible or hidden? Is every dependency indicated in both directions? Does a change in one area of the notation lead to unexpected consequences?}

\DSFs{} do not themselves have explicit notion of dependencies. Instead, their relationship with this dimension relates primarily to their usage. For instance, in maintaining a package.json for a JavaScript project an alteration to the version of one library might cause a huge number of additional libraries to be installed (as implied dependencies), which may in turn cause version conflicts with other explicitly installed packages.
Beyond usage, the primary area of \cdn{hidden dependencies} is in the schema, which in some usages is essentially a part of the notation itself.
One practitioner noted: \hn{Someone would change the code and always forget to update the schema to match.}~\cite{galonk}

\parahead{\cdn{Viscosity}} \emph{Are there any inherent barriers to change in the notation? How much effort is required to make a change to a program expressed in the notation?}

From a comparative notational perspective, how strict a notation is directly informs how viscous it is. For instance, JSON strictly requires commas between list items and not for the final item (\eg{} \verb+[1, 2, 3]+ is valid while \verb+[1, 2, 3,]+ is not).
Addressing these syntactic variations are the primary differences between formats. For instance, HJSON and JSON5 make those quotes optional and allow trailing final commas.
In contrast, XML increases viscosity by requiring closing tags.

In addition to basic syntactic rules such as commas and tags, viscosity is also driven by structural characteristics of the notation. Formats with nested hierarchies amplify the effort of even localized edits, as changes may cascade across multiple levels. YAML’s reliance on indentation makes it especially fragile, where small whitespace shifts can break an entire file.
More broadly, in depends on the scope of the change. CSV exhibits a distinct profile due to its flat structure: while row-level modifications are straightforward, column-level changes---such as reordering fields---might require global updates.
Updating a value is typically a localized, isolated change for all \DSFs{} we examine. However, field-level changes often require multiple coordinated edits throughout the file to ensure consistency.
Among the formats examined, only CSV largely avoids this problem, since in flat tabular data, field names usually appear only once in the header row. While Other formats, however, require global updates wherever the field name is referenced, resulting in high viscosity. This challenge is frequently discussed in developer communities,
for instance users express frustration over the repetitive nature of bulk modifications: \hn{The JSON format is just too verbose for me - requiring the repeated "column" tag when the JSON itself handles arrays of values is troublesome.}~\cite{Guest}
The modest variable like abstractions in XML and YAML offer limited support for reuse, but their benefit in reducing viscosity is marginal unless the modification precisely targets a reused structure.

\parahead{\cdn{Closeness of Mapping} and \cdn{Role-Expressiveness}} \emph{How closely does the notation correspond to the problem domain world?} and  \emph{How obvious is the role of each component of the notation in the solution as a whole? }

\DSFs{} are based around a small collection of data structures, roughly records and arrays, and closely map to how a variety of different languages model those data structures. For instance JSON directly models the notions of objects present in JavaScript, hence the name.
As one user on Reddit noted, \hn{
    JSON is a text format. That’s all. It’s a specific way of formatting text following a well defined structure to organize data and assign meaning to it.}~\cite{jtf} In JSON, curly braces ``\{\}'' denote objects, square brackets ``[]'' represent arrays, and colons ``:'' connect keys to their associated values. While these conventions must be learned, they are stable, consistent, allowing users to readily understand and distinguish the function of structural components during both reading and writing.
Beyond these minor syntactic structural differences, the \cdn{closeness} and \cdn{expressiveness} depends on the context in which it is used. For instance, Vega~\cite{satyanarayan2015reactive} (a DSL based in JSON) has a very close mapping to the structure of SVG, which is the basis of its rendering model.
Overall, there are no substantial differences between the formats.

\parahead{\cdn{Consistency}} \emph{After part of the notation has been learned, how much of the rest can be successfully guessed---either by recombining known elements or by applying learned rules to unfamiliar ones?}

Most data serialization formats we considered---such as CSV, JSON, XML, and TOML---pursue internal consistency through a strict, uniform, and small syntax. This design promotes reliable parsing and enables users to develop stable expectations about how the format behaves in new situations.
While many formats strive for consistency in principle, practical factors such as syntactic flexibility, version drift, and incompatible tooling ecosystems can substantially erode a user’s ability to extrapolate from known patterns. YAML, for instance,  permits unquoted strings and includes some special keywords which are mapped to specific values (for instance in YAML 1.1 ``yes'' is interpreted as a boolean and not a string).
The same data can be represented in a variety of stylistic variations across projects or contributors.
JSON5 and HJSON similarly relax standard JSON’s constraints by allowing trailing commas, comments, unquoted keys, and even omitting commas entirely in some cases.
While these changes improve editability and reduce common syntax errors, they also make it harder to guess the correct form based on previously learned patterns. TOML generally maintains uniformity but still exhibits minor quirks, particularly in specialized domains such as date and time handling.

\parahead{\cdn{Error-Proneness}} \emph{To what extent does the notation influence the likelihood of user errors?}

In general, stricter formats tend to reduce ambiguity but increase the cost of small mistakes, while more flexible ones may trade off strict validation for improved tolerance---often at the risk of introducing less obvious errors. Formats like JSON and XML offer clear and rigid structure, but minor issues such as a missing comma, unmatched bracket, or misplaced character can invalidate the entire document. Such brittleness is frequently reflected in community forums, where users seek help diagnosing syntax problems: \hn{The JSON you've posted is invalid---not because of the trailing comma at the end, but due to a missing comma in the center.}~\cite{ChuckConway} Several JSON variants (\eg{} JSON5, HJSON, and JSONC) relax these rules, supporting comments, trailing commas, and unquoted keys---their error-proneness differs depending on the degree of tolerance. CSV often has errors that often arise from misplaced commas that shift fields and break parsing.
YAML, on the other hand, is particularly susceptible to invisible or misleading errors due to its sensitivity to indentation. A single misplaced space can change the structure in subtle, hard-to-detect ways. As a Reddit user complained: \hn{I ended up spending so much time trying to figure out why the command didn't work only to find out it was spacing that was screwing it up.}~\cite{kyledreamboat}
In addition, YAML underwent significant semantic changes between versions 1.1 and 1.2: values like yes, on, or 22:22 are interpreted as booleans or sexagesimal numbers in YAML 1.1 but treated as strings in YAML 1.2 which results in \hn{Yaml 1.2 differs substantially from 1.1: the same document can parse differently under different YAML versions.}~\cite{RuudvanAsseldonk}
Users unaware of this change might rely on these shortcuts.
Complicating this matter further is that some 1.2 parsers accept the 1.1 version of booleans.

\parahead{\cdn{Hard Mental Operations}} \emph{How much hard mental processing lies at the notational level, rather than at the semantic level? Are there places where the user needs to resort to fingers or penciled annotation to keep track of what's happening?}

Here, hard mental operations stem from the need to recall and apply syntactic rules---such as the roles of colons, quotation marks, and nested brackets in JSON, or the requirement to correctly pair opening and closing tags in XML.
These demands can be taxing in formats with deep nesting or verbose markup, where users must mentally track context, delimiters, and hierarchical scope across many lines of code---linking this difficulty of the operations to the syntactic viscosity.
In contrast, one user praised YAML for its relatively lower syntactic churn, observing that \hn{I don't have to put everything in quotes, then in braces and brackets.}~\cite{RollForPerception} Yet YAML also has its own challenges: indentation sensitivity makes structure fragile, while implicit typing requires users to anticipate how values will be interpreted (\eg{} ``yes'' as booleans). In CSV, while its flat structure avoids hierarchical nesting, users must rely heavily on header rows to interpret content, and wide tables make columns difficult to differentiate.

Notably this dimension is significantly alleviated by the surrounding tool ecosystem which includes syntax highlighting, automatic bracket matching, and code folding.
However, the necessity of these aids underscores the mental cost of working with structurally rigid notations without visual scaffolding.

\parahead{\cdn{Diffuseness / Terseness}} \emph{How many symbols or how much space does the notation require to produce a certain result or express a meaning?}

A key point of disagreement in preferences about \DSFs{} is what amount of syntactic verbosity, here modeled as symbolic diffusion is acceptable.
Formats like XML are typically considered highly diffuse due to their reliance on extensive structural markup, including explicit opening and closing tags (\eg{} \texttt{<item></item>})---however, arguably this builds in a degree of error correction by driving the user to match the tags.
JSON by contrast is substantially less verbose, providing a relatively compact way to represent objects and arrays but still requires structural punctuation such as braces, brackets, and quotation marks.
JSON5, HJSON are terser still, allowing for more compact representations by omitting quotes, commas, and sometimes even braces.
YAML minimizes the amount of structural characters and allows the data to show itself in a natural and meaningful way~\cite{yamlSpec} by centering whitespace, however YAML also includes several features not present in other formats.
In formulating the StrictYAML parser, O'Connor~\cite{strictyaml-why-not-toml} specifically removed a number of the features including implicit typing, binary data and explicit tags in to conceptually simplify the usability of the format. YAML and XML's support for variables also increases their diffuseness. CSV avoids structural verbosity altogether through a flat single-field delimiter scheme, though this simplicity constrains its expressive power.

\parahead{\cdn{Juxtaposability}} \emph{Can different parts of the notation be compared side by side at the same time? How easy with which can users visually compare related pieces of data?}

As datasets grow larger---with more fields and deeper nesting---differences across data serialization formats become more pronounced in this dimension.
CSV offers the strongest juxtaposability, allowing values in the same column to be easily compared across rows. At the other end of the spectrum, XML performs the worst. Its verbose syntax, especially the use of explicit opening and closing tags (\eg{} \texttt{<field></field>}), introduces significant visual clutter, forcing users to scroll extensively just to compare similar fields.
Formats not based on white space (\eg{} JSON) can remove line breaks to increase \cdn{juxtaposability}, but the extent to which this is effective is mediated by screen size and text length.
To that end, all of these are strongly effected by length of the input by relative factors of syntactic terseness.
This limitation is closely tied to the lack of abstraction available across formats. XML and YAML provide partial affordances for abstraction (\eg{} variables), offering at least some capacity to mitigate the challenges associated with this dimension.

\parahead{\cdn{Visibility}} \emph{How readily can required parts of the notation be identified, accessed and made visible? }

Visibility refers to how easily users can locate and identify relevant information within a notation. Across common data serialization formats like JSON, YAML, XML, TOML, and CSV, the difference in visibility is often not pronounced.
Difficulties in locating specific values are more often a result of data size or structural depth, highlighting the impediments caused by a limited level of abstraction.
XML, for example, can create a higher visual load due to verbose tag structures and repeated element names, which can visually obscure meaningful data by pushing it deeper into the document.
Similarly, JSON’s use of brackets and nesting can add bulk, but the structure remains logically transparent. In practice, users often rely on tools like Ctrl+F to search for keywords, regardless of the format. While deep nesting or long documents may challenge visibility, no format inherently renders information hidden or un-discoverable, merely adding friction to the process.

\parahead{\cdn{Premature Commitment}} \emph{Are there strong constraints on the order in which the user must complete the tasks to use the system?}

Most data serialization formats support incremental construction of data structures without requiring the user to declare the full structure in advance, although there are limitations to this.
In particular hierarchical data must be specified in some order, typically by defining the parent container before specifying its children.
However some structural features have less strict ordering. for instance TOML's section headers (\eg{} [profile], [profile.address]) can be added before or after the contents of the section are written.
In addition to structural constraints, the level of strictness in type specification can also lead to premature commitment. JSON requires that strings be enclosed in double quotes while numbers must not, forcing users to decide on data types early in the writing process. By contrast, formats such as YAML, TOML, and XML allow bare strings, reducing the need for early type declarations. When some formats combined with schema systems (\eg{} JSON Schema or XML Schema Definition), or when formats such as YAML or XML require variables to be defined before use, further increasing the demand for upfront decisions.

\parahead{\cdn{Progressive Evaluation}} \emph{How easy is it to evaluate and obtain feedback on an incomplete solution?}

Progressive Evaluation is not an inherent property of the data serialization format itself, but is largely shaped by the capabilities of the editing environment. Whether users can preview or validate partial or incomplete structures depends less on the syntax or strictness of the notation, and more on the presence of tooling features such as real-time linting, auto-completion, suggestions, or partial rendering. In modern development environments like VS Code, these features are widely available across most formats. As a result, the extent to which a format ``supports'' progressive evaluation has limited practical impact---tooling tends to flatten the differences. Notably, formats that are used in conjunction with schema systems offer enhanced error detection and explanatory feedback, which can support more effective progressive evaluation.

\parahead{\cdn{Secondary Notation and Escape from Formalism}}
\emph{Can the notation carry extra information by means not related to syntax, such as layout, color, or other cues?}

For these formats, extra information involves the addition of comments.
While some common formats (CSV and JSON) do not support comments directly (as they only support comments through undesigned for means~\cite{jsonComments}), most other formats include some commenting mechanism.
While simplistic, this support allows users to embed contextual explanations without altering the underlying data semantics.

\subsection{Data Serialization Formats Usability Interview Guide}
\label{appendix:interview-guide}

\subsection*{Introduction}

\noindent
Hello, thank you very much for participating in this interview. My name is REDACTED, and this interview is aimed at understanding the usability of data serialization formats. Your responses will be kept completely confidential and used solely for research purposes. The interview is expected to take approximately 30 minutes. \par
We will record the interview process for data analysis and anonymize your personal information. Your participation is completely voluntary; you may stop or skip any questions at any time.

\subsection*{Consent Statement}

\noindent
Before we begin, please review the consent statement to understand your rights and how your data will be used. If you have any questions, feel free to ask at any time. Once you have confirmed your agreement, please complete and submit the Google Form. \par
Now, we will begin the recording.

\subsection*{Interview Questions}

\begin{enumerate}
    \item Which data serialization formats do you commonly use in your work? Could you specify them with their usage (\eg{} JSON, YAML)?
    \item Do you have to maintain configuration or data files as part of your day-to-day work?
          \begin{itemize}
              \item If so, which serialization formats do you use?
              \item What actions do you usually take with them?
              \item How well do you feel these formats support your frequent maintenance needs?
              \item Are there any aspects you find particularly satisfying or frustrating?
              \item Do you believe other formats could perform better in this regard? Why?
          \end{itemize}
    \item Do you maintain data/config files less frequently (\eg{} monthly or long-term)?
          \begin{itemize}
              \item If so, which serialization formats do you use?
              \item What actions do you usually take with them?
              \item How well do you feel these formats support your frequent maintenance needs?
              \item Are there any aspects you find particularly satisfying or frustrating?
              \item Do you believe other formats could perform better in this regard? Why?
          \end{itemize}
    \item How do you decide which formats to use in different projects?
    \item What are your primary programming languages and tech stack? Do they influence format choice?
    \item Do you prefer using a single format or multiple based on task-specific needs?
    \item Please share your impressions of the following formats based on your own experience or community discussions:
          \begin{itemize}
              \item JSON, JSONC, JSON5, HJSON, YAML, TOML, XML, CSV
          \end{itemize}
    \item Among the formats, which one supports searching/modifying best in your workflow?
    \item Which are best for collaboration and reuse? Why?
    \item Which do you tend to avoid? Why?
    \item Which have been most frustrating to work with? How did that affect productivity?
    \item What debugging/error handling issues have you encountered?
    \item When handling large data, which formats perform best? Why?
    \item Which formats are most user-friendly? Why?
    \item What future features would improve your experience with \DSFs{}?
\end{enumerate}

\subsection*{Wrap Up}

\noindent
Is there anything else you'd like to share before we conclude the interview? \par
We will now stop the recording. Thank you again for your participation! If you have any further questions about this interview, feel free to contact us.

\subsection*{Post-Interview Survey}

\noindent
After the interview, we will send a short post-interview survey (within one day) to collect demographic information and contact details for a digital gift certificate. Please complete and submit it upon receipt.

\end{document}